# Identifying candidate hosts for quantum defects via data mining

Austin M. Ferrenti,[†] Nathalie P. de Leon,[‡] Jeff D. Thompson,[‡] and R. J. Cava[†]

† *Department of Chemistry, Princeton University, Princeton, NJ 08544, USA*

‡ *Department of Electrical Engineering, Princeton University, Princeton, NJ 08544, USA*

**Atom-like defects in solid-state hosts are promising candidates for the development of quantum information systems, but despite their importance, the host substrate/defect combinations currently under study have almost exclusively been found serendipitously. Here we systematically evaluate the suitability of host materials by applying a combined four-stage data mining and manual screening process to all entries in the Materials Project database, with literature-based experimental confirmation of band gap values. We identify 580 viable host substrates for quantum defect introduction and use in quantum information systems. While this constitutes a significant increase in the number of known and potentially viable material systems, it nonetheless represents a significant (99.54%) reduction from the total number of known inorganic phases, and the application of additional selection criteria for specific applications will reduce their number even further. The screening principles outlined may easily be applied to previously unrealized phases and other technologically important materials systems.**

# Introduction

In recent years, significant effort has been devoted to the realization of functional systems for quantum information science (QIS). QIS devices employ the manipulation of quantum states to store, process, and transmit information, potentially enabling the rapid solution of problems long thought impossible or impractical to address through classical methods. One of the most promising platforms for the development of efficient quantum information systems, particularly for application in quantum networks,[1] is atomic or atom-like defects in solid-state hosts.[2] The quantum coherence characteristics of the atomic defect, low defect concentrations, and quality of the host when combined should allow for long spin coherence and efficient optical transitions, properties also ideal for nanoscale sensing under ambient conditions.[3]

Several such atomic defect systems, including vacancy centers in diamond[4–9] and silicon carbide[10] have been extensively studied for several applications, in particular quantum networks,[11,12] magnetometry and nanoscale sensors of magnetic fields,[3,13,14] electric fields,[15] temperature[16,17] or chemical composition using NMR,[18,19] often under ambient conditions at room temperature.[20] Another well-known class of defects is transition metal and rare earth ion impurities. These defects have been extensively explored in the context of solid-state laser development, in materials where extremely high doping concentrations are possible. However, exploring single defects as qubits is a more recent development,[21,22] partially facilitated by integration with nano-photonic circuits to modify and enhance their luminescence.[23,24] Although significant attention has thus far been paid to a handful of materials and defects, they represent only a small fraction of the greater body of potential defect-host systems. As most known quantum information materials systems have been discovered indirectly, a rational search of inorganic materials can yield candidates with superior properties.

Recent work has focused on *ab initio* predictions of host-defect systems that can be carried out with electronic structure calculations, as recently demonstrated with vacancy centers in silicon carbide,[25] and such calculations have been carried out for several individual host-defect combinations. An alternative approach is to systematically screen for potential host materials by first postulating which properties would be ideal for host material candidates, and then screen materials based on those properties. This is done most efficiently through a computational search that substantially narrows the field of potential candidates.

In order to achieve the long spin coherence and high efficiency optical transitions necessary for a usable quantum information system, host substrates must be highly pure (i.e. as free from defects as possible), intrinsically diamagnetic (thereby reducing magnetic noise and spin-based relaxation of the defect quantum state), and possess a band gap large enough to accommodate the ground and excited energy levels of the defect (separated by an optical frequency). The ideal, pure substrate would be free of paramagnetic impurities or unwanted defects that may influence the band gap character and/or a magnetic or electric field environment surrounding the implanted defect, reducing efficiency. Those materials known to be dopable, to possess controllable surfaces, and to possess a known method of epitaxial thin-film production would further facilitate the production of integrated devices.

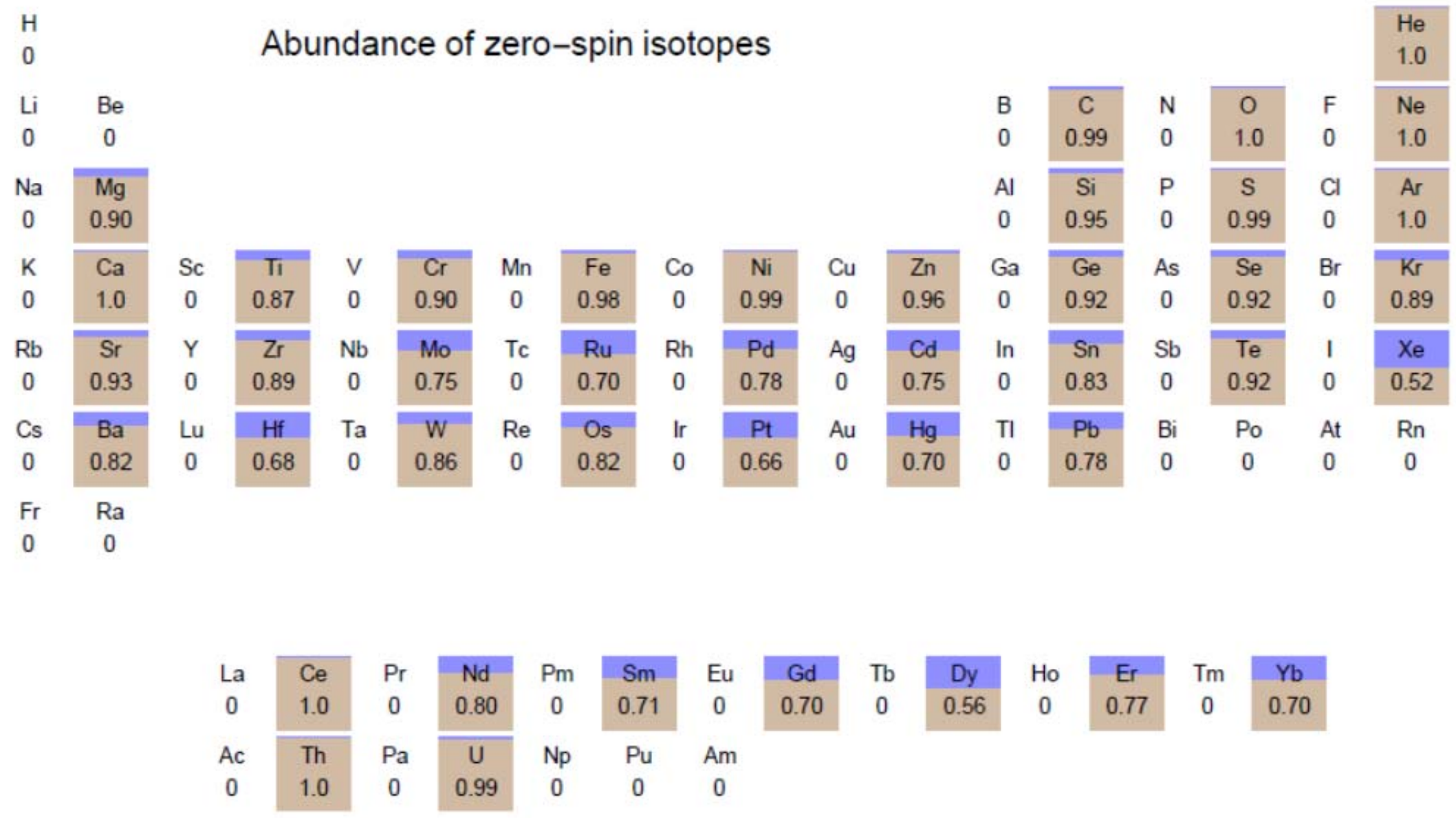

**Figure 1. Color-coded periodic table depicting the natural abundance of stable zero nuclear spin isotopes,** represented by the portion shaded gold ($I$=0) relative to the portion shaded blue ($I \neq 0$). Those elements for which there exist no stable spin-zero isotopes are shaded white.

Reducing magnetic noise and relaxation in the environment of the defect within the host requires both the minimization of paramagnetic centers in the host and the absence of host nuclei with non-zero magnetic moments. Consideration of the natural abundance of zero nuclear spin isotopes of each element, shown in Figure 1, allows for the exclusion of a significant portion of the periodic table, for which no stable spin-zero isotopes exist. While some elements, such as O and Ca, are known to exist almost exclusively as nuclear-spin-free isotopes, elemental species with at least 50% nuclear-spin-free isotopes could likely be isotopically enriched to higher concentrations, as has been achieved in diamond[26,27] and silicon[28]. Although several of the lanthanide elements appear to have relatively high natural abundances of spin-free isotopes, the difficulty in obtaining pure, diamagnetic starting material of these (free from magnetic lanthanide impurities) excludes them from consideration here. All transition metal elements with unpaired electrons are eliminated due to their paramagnetism. As the optical coherence of defects may be sensitive to the presence of permanent electric dipole moments, and thus also to their site symmetry, phases crystallizing in polar space groups were also not considered.

In the current work, a systematic application of each of the criteria outlined above was performed, beginning with the totality of known inorganic phases listed in the Materials Project database. As the practice of database mining is often highly automated, the sensitivity of the desired properties to the exact phase and crystal structure reported for each potential host species necessitated manual checks after each restriction of the data set. While the true number of viable quantum host-defect pairings may be effectively infinite due to the fact that different quantum defects will be suitable for different applications, of the 100,000+ inorganic materials in the inorganic crystal structure database (ICSD), a maximum of 580 phases were found to be potentially viable hosts for atom-like defect quantum information systems. Among these, a small number are simple single elements or binary compounds, making them relatively straightforward to prepare and study as substrates for QIS.

# Results and Discussion

To reduce the possibility of accidental exclusion of viable host materials, the screening of the Materials Project's 125,223 inorganic compound entries was conducted in four distinct stages, as shown in Figure 2.

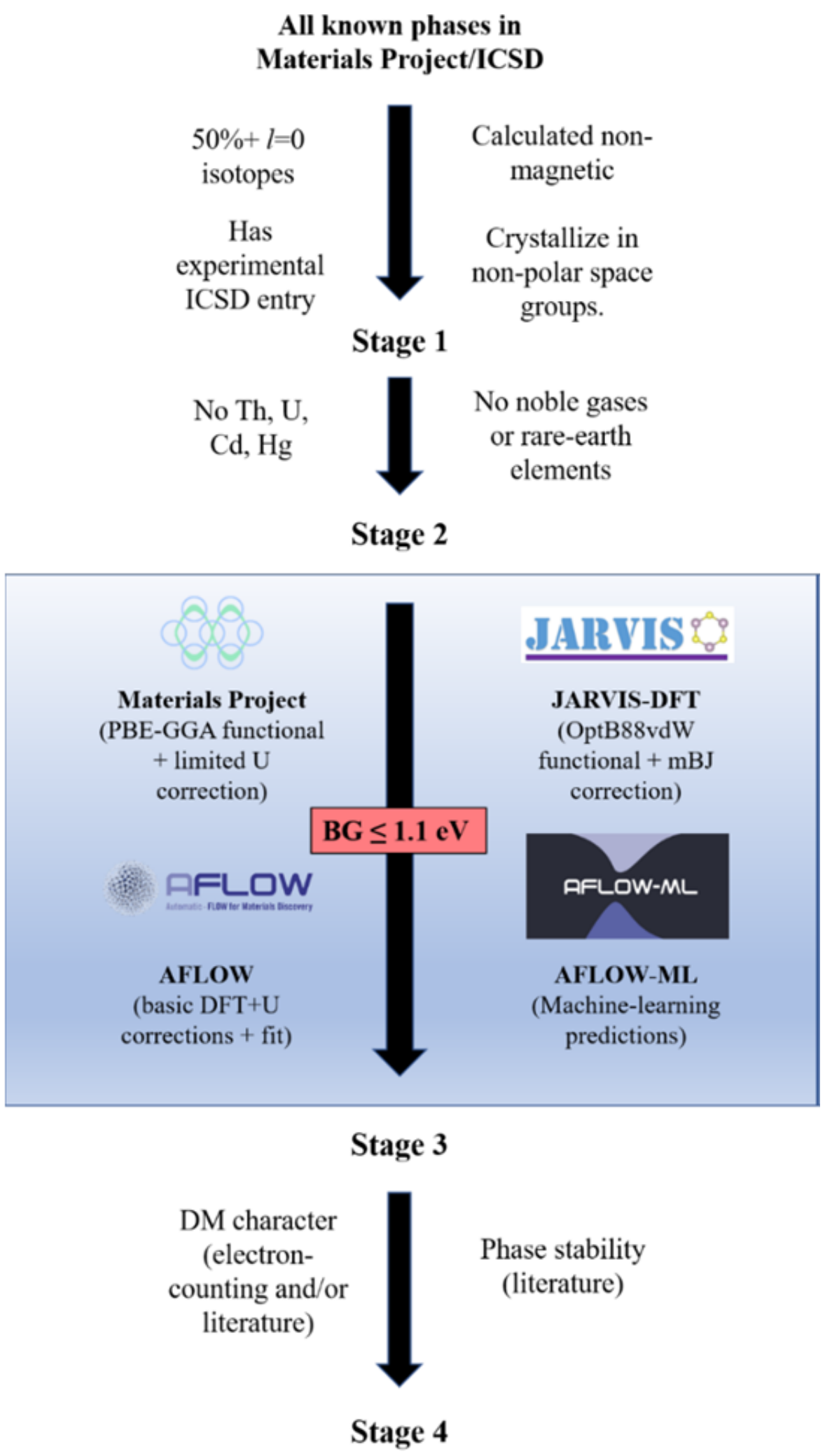

**Figure 2**. **Four-tier screening process to evaluate the potential viability of materials as hosts for quantum defects.** Stage 1) Beginning with all known experimental inorganic phases, remove those with calculated non-zero net magnetic moment, crystallization in polar space groups and/or containing atomic species with <50% $I$=0 isotopes. Stage 2) phases containing radioactive Th and U, toxic Cd and Hg, and the often magnetically impure rare earth elements were removed. The one phase containing a noble gas element that passed this sieve, orthorhombic $XeO_3$, was also removed due to its known instability under standard conditions. Stage 3) The calculated, predicted, and measured experimental band gaps of the remaining phases were recorded from the Materials Project, the JARVIS-DFT database, the AFLOW repository, the AFLOW-ML API, and the existing literature. Those materials that could not be reasonably predicted to have a band gap larger than 1.1 eV were dismissed. Stage 4) The intrinsic diamagnetic (DM) character and phase stability under standard conditions of the remaining candidates were confirmed using a combination of standard electron-counting principles and literature reports. All database searches were conducted in February and March of 2020.

1) The first screen was applied utilizing the Materials Project's built-in query API[29], and removed all entries corresponding to phases crystallizing in polar space groups, derived from an experimental ICSD entry, containing only those elements with a greater than 50 percent natural abundance of zero nuclear spin isotopes, and calculated to be non-magnetic systems.

2) The second screen was applied by hand, and included the removal of all phases containing uranium, thorium, cadmium, and mercury, due to the radioactive and/or toxic nature of the majority of stable phases that each may form, as well as those containing noble gases, as none exist as stable solids under standard conditions. All rare-earth-element-containing phases were also removed, due to the difficulty in obtaining sufficiently pure starting materials free of nuclear spin.

3) The third and most important filter was applied utilizing calculated band gaps for all remaining phases reported by three publicly-available DFT computation repositories; the Materials Project, the Joint Automated Repository for Various Integrated Simulations (JARVIS), and the Automatic Flow of Materials Discovery (AFLOW) library, each of which employs a different band gap calculation and/or prediction method. In addition, where available, the Machine Learning-predicted band gap reported by Isayev *et al* in their predictive modeling studies was also taken into consideration.

At this stage, those phases with a Materials Project band gap of 2.0 eV were deemed likely to possess a large enough band gap to accommodate the necessary optical transitions of useful quantum defects, and, so as not to miss any potentially viable candidates, only those with calculated band gaps <0.5 eV were removed. For all other phases, any reports of experimental or otherwise calculated band gaps in the literature were taken into consideration. Various higher-order compounds for which the calculated band gaps were insufficient, such as in the Ba-Hf-S family, were approved based on general band gap trends. Those phases with calculated and/or experimentally-derived band gaps consistently in the range of that of Si, which would represent the lower bound of viability for potential host materials, are grayed in Tables 1-4.

**The Materials Project**[30] utilizes the DFT-based Vienna ab initio simulation package (VASP) software[31] to calculate a wide variety of structural, energetic, and electronic properties for all reported inorganic structures in the Inorganic Crystal Structure Database (ICSD). An initial relaxation of cell and lattice parameters is performed using a 1,000 *k*-point mesh to ensure that all properties calculated are representative of the idealized unit cell for each material in its respective crystal structure. To calculate band structures for these materials, the Generalized Gradient Approximation (GGA) functional is applied to the relaxed structures. For structures containing one of several transition metal elements such as Cr, Fe, Mo, and W, the +U correction is also applied to correct for correlation effects in occupied d- and f-orbital systems that are not addressed by pure GGA calculations.[32] The authors caution that due to the high computational cost of more sophisticated calculation methods, those employed often produce severely underestimated band gaps relative to the experimentally-derived values.

**JARVIS-DFT**[33], originally compiled as a database for functional materials, with a focus on the discovery of novel two-dimensional systems, also employs the DFT-based VASP software to perform a variety of material property calculations. As opposed to the Materials Project, JARVIS-DFT employs the OptB88vdW (OPT) functional, which was initially designed to better approximate the properties of two-dimensional van der Waals materials, and has since also been shown to be effective for bulk systems.[34,35] Structures are first sourced from the Materials Project database, and then re-optimized using the OPT functional. A representative band gap is then calculated through a combination of the OPT and modified Becke-Johnson (mBJ) functionals. The mBJ and combined OPT-mBJ functionals have both been shown to predict band gap sizes with more accuracy than other DFT-based calculation methods.[36]

The **AFLOW**[37] repository relies on a highly-sophisticated and automatic framework for the calculation of a wide array of inorganic material properties.[38] The GGA-based Perdew–Burke-Ernzerhof (PBE) functional with projector-augmented wavefunction (PAW) potentials is first used within the VASP software to twice relax and optimize the ICSD-sourced structure using a 3,000-6,000 *k*-point mesh. The increased *k*-point mesh density, compared to that employed by the Materials Project, is indicative of a more computationally-expensive calculation. The band structure is then calculated with an even higher-density *k*-point mesh, as well as with the +U correction term for most occupied d- and f-orbital systems, and the standard band gap ($E_{gap}$) is extracted.[39] A "fitted" band gap ($E_{gfit}$) is then calculated by applying a standard fit, derived from a selective study of DFT-computed vs. experimentally-measured band gap widths, to the initially calculated value.[40]

**AFLOW-ML**[41], a machine learning API designed to predict thermomechanical and electronic properties based on chemical composition alone, further builds upon the entries present in the ICSD and calculated through the AFLOW framework. Using only provided atomic compositional and positional information, so-called "fragment descriptors", the system first applies a binary metal/insulator classification model. For materials predicted to be insulators, an additional regression model is applied to predict the band gap width. Each model was subjected to a fivefold cross validation process, in which it was trained to more accurately predict properties in an independent data set. The initial binary classification model is shown to have a 93% prediction success rate, with the majority of misclassified materials being narrow-gap semiconductors. While the accuracy of the predicted band gap sizes relative to experimental values is not mentioned by the authors, roughly 93% of the machine learning-derived values are found to be within 25% of the DFT+U-calculated gap width. Only those phases identified in the authors' initial cross-validated test set were used for comparison.

4) The final stage. The criteria by which potential host species were excluded in the first three stages were largely derived computationally, with subsequent manual checks. In contrast, the final stage of screening involved the confirmation of various of the fundamental properties of the materials found in the existing literature.

First, for those remaining stoichiometries for which multiple phases appeared to be viable, the relative stability of each polymorph at STP was recorded. All recognized high-pressure and high-temperature phases that were not reported to be quenchable to a stable state under standard conditions were removed.

The intrinsic magnetic character of each pure phase was then confirmed through a combination of standard electron-counting rules and literature reports. As many of the potentially viable materials contained oxygen, particular care was taken to consider whether reported paramagnetic character could be due to oxygen-defects, especially in the various molybdate, platinate, and palladate phases. Any pure phase that was reported to deviate from diamagnetic character was removed.

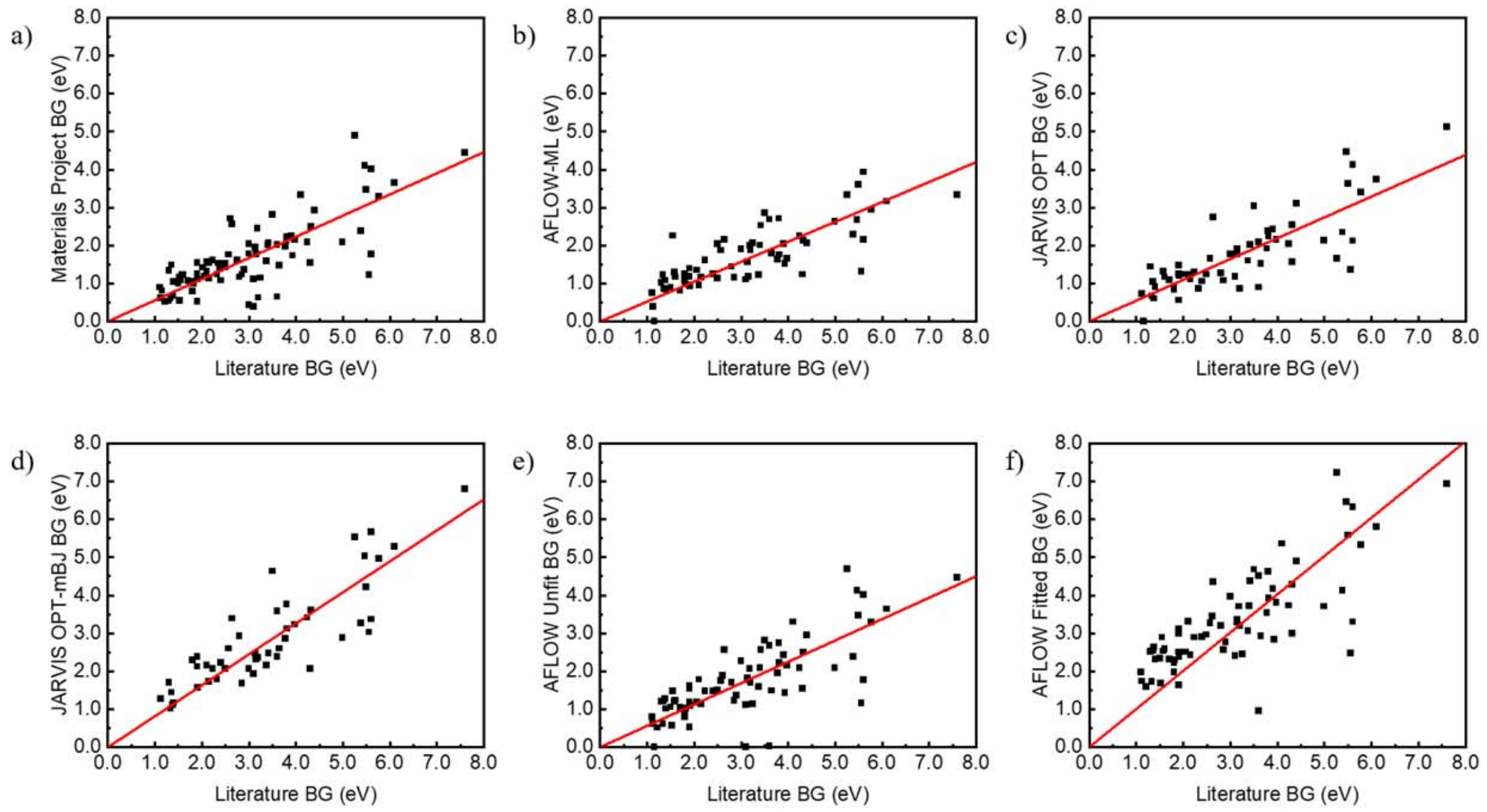

**Figure 3. Comparison of reported experimental band gaps to those calculated** by a) the Materials Project b) AFLOW-ML c) JARVIS-DFT (OPT only) d) JARVIS-DFT (OPT-mBJ) e) AFLOW (unfit) f) AFLOW (fit). On average, the standard DFT calculations (a, c, and e) and machine-learning predictions (b) were found to underestimate the experimental band gaps by roughly 40-50%. JARVIS-DFT OPT-mBJ calculations (d) were found to underestimate measured values by about 18% on average, while AFLOW fitted band gap calculations (f) were found to overestimate by about 2% on average, with several significant outlying underestimations. The red lines show the linear least squares fits to the data.

Where the true size of a material's band gap could not be reasonably assumed suitable based on calculations alone, other reported computationally and experimentally-derived band gap values were also taken into consideration when available. While experimental band gaps were either not available or not recorded for the majority of phases considered, those that were available were compared to the calculated values from each database, in order to better judge the viability of the selected materials with small calculated gaps (Figure 3). The standard band gap calculations performed by the Materials Project, JARVIS-DFT, and AFLOW databases were found to be underestimated by roughly 40-50%, relative to measured values, consistent with the long-known inaccuracies of DFT estimates.[42] AFLOW-ML's machine learning-based predictions were underestimated to a similar degree. On average, the OPT-mBJ hybrid functional employed by the JARVIS-DFT database for some phases was found to reduce this underestimation to 18.4%, while the "fitted" band gap calculated by AFLOW was actually observed to be overestimated by about 1.8% relative to experimental values. However, the latter figure is heavily influenced by several outlying underestimations in the AFLOW “fitted” gap data set, suggesting that on average, band gaps calculated in this manner will be overestimated to a greater degree. The determination of band gap suitability was thus made with these findings in mind.

It should be noted that despite appearing in the literature, the most stable phases of several, long-recognized materials were found to lack an experimental crystal structure in the ICSD, and subsequently were also absent from the Materials Project database. Several of these, such as the STP-stable phases of $BaGeO_3$, $BaGe_2O_5$, and $C_{70}$ would be potentially viable host species, but due to their absence from any of the databases studied are not included. However, while comparing computed band gaps with reported literature values, three additional phases were identified that lacked a corresponding Materials Project entry, but did appear in at least one of the other databases. fcc-$C_{60}$ is the only included phase with a corresponding ICSD entry that did not appear in any of the databases considered.

While suitably-sized single crystals are necessary for the fabrication of functional devices for quantum information systems, few materials have been reported to be easily grown as large, defect-free, and optically clear single crystals. As such, the existence of published single crystal synthesis, regardless of product size, is denoted in the "SC" column in the tables of results by an asterisk. Noted air and moisture instability in the literature was also considered, but was not exclusionary, as additional post-synthesis processing may allow otherwise viable host materials to be utilized.

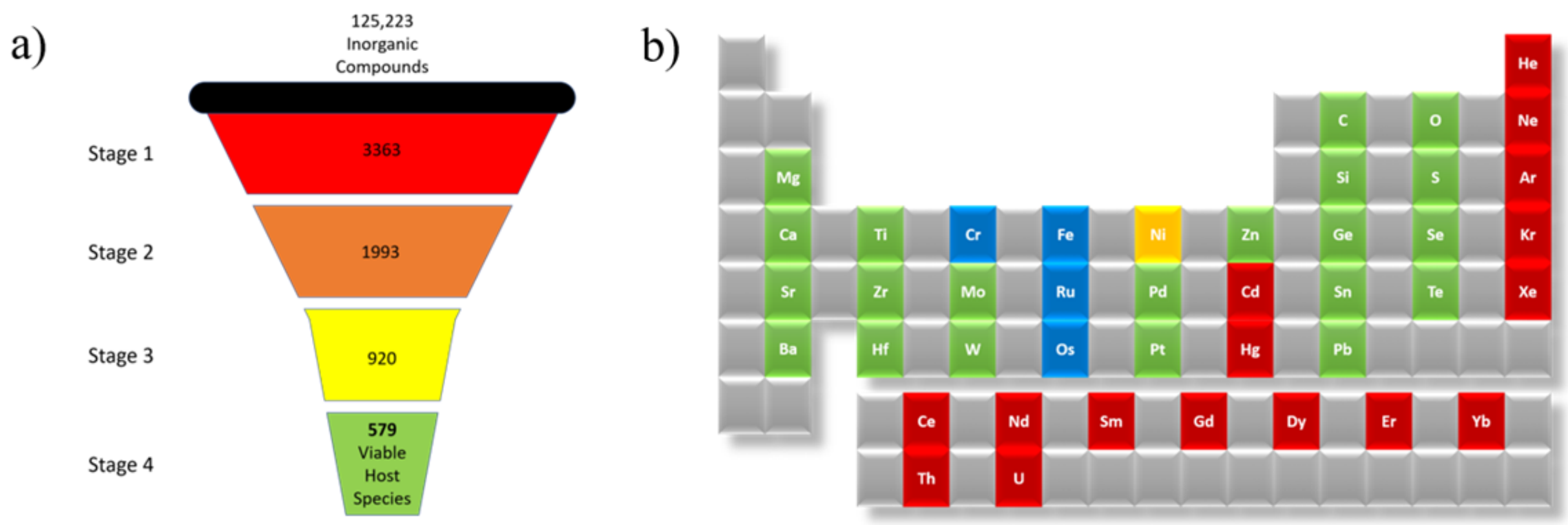

**Figure 4**. **Tiered screening system for the identification of potentially viable host species for quantum defect implantation.** a) A schematic view of the process. A maximum of 0.46% of all phases in the International Crystal Structure Database were found to be potentially viable host substrates, with the most substantial reduction in candidate phases (97.3%) occurring in Stage 1 of the screening process, b) Color-coded periodic table of constituent elements in viable host species depicting elements: disregarded due to the absence of spin-zero isotopes (gray); deemed too hazardous, too radioactive, too difficult to purify, and too unreactive for inclusion (red); not present in any identified phases (yellow); present largely only in cluster complex phases (blue); and present in numerous identified phases (green).

Of the 125,223 inorganic compounds listed in the Materials Project database, a maximum of 580 phases (0.46%) are found to fit the criteria outlined above as suitable hosts for quantum defects (Figure 4a). While there likely exist significantly more potentially viable phases that have yet to be recognized, this total spans all currently reported experimentally synthesized and theoretically predicted inorganic materials. As the ease of growth of suitably-sized single crystals, as well as stability under standard or near-standard atmospheric conditions, and chemical and structural simplicity are major factors in determining the viability of potential host phases, it is likely that those identified here represent the majority of suitable candidates.

Of the 580 identified phases, the band gap character of at least 560 could be either confirmed or reasonably assumed to be comparable or larger than that of silicon (≥1.1 eV), based on calculated and/or reported experimental values. The band gaps of the remaining 20 candidates, grayed in Tables 1-4, have either been reported to be slightly smaller than that of Si, or could only be tentatively extrapolated to a similar value.

Only one admitted element, Ni, was not found to be present in any of the listed phases, as all Ni-containing materials considered were found to be intrinsically paramagnetic. All the Cr-,

Fe-, and Os- containing species in the tabulations, as well as the majority of those containing Ru, are transition-metal based carbonyl cluster compounds.

# Conclusion

From the total number of known, experimentally synthesized inorganic material phases, we have derived a catalog of all potentially viable host substrates for functional, atom-like defect-based quantum information systems. A combination of automatic database and subsequent manual screening was performed to ensure that each reported phase possessed the necessary electronic, magnetic, and optical properties, as well as stability under standard conditions, necessary for application in QIS. This screening was conducted across several materials databases (including the Materials Project, JARVIS-DFT, and the AFLOW repository), the AFLOW-ML prediction API, and the existing body of literature in an attempt to compensate for known computational and experimental biases and errors that may have influenced our results. Of the 580 viable candidates found, 17 are unary, 77 are binary, 337 are ternary and 149 are quaternary or higher. Many are oxides and chalcogenides, suggesting that they will be relatively simple to fabricate and study, while others in the tabulation, such as the osmates and higher order elemental clusters, will likely prove impractical for currently considered device applications. Future studies concerning inorganic material systems would likely also benefit from the implementation of similar systematic screening processes, enabling the rapid and conclusive determination of all viable phases for any number of applications.

# Acknowledgements

Funding for this research was provided by the AFOSR (contract FA9550-18-1-0334), the Eric and Wendy Schmidt Transformative Technology Fund at Princeton University and the Princeton Catalysis Initiative. The authors acknowledge helpful conversations with Chris Phenicie, Paul Stevenson and Sacha Welinski.

# Competing Interests

The authors declare no competing financial or non-financial interests.

# Data Availability

The authors declare that the data supporting the findings of this study are available within the paper.

# Author Contributions

J.D.T., N.P.dL., and R.J.C devised the main conceptual ideas and guiding principles for the project. A.M.F. performed the data screening and analysis. R.J.C. supervised the research. A.M.F. wrote the manuscript with input from all authors.

**Table 1. The potentially viable unary host materials.** All band gaps are reported in eV. Phases reported in the JARVIS-DFT database for which both standard OPT and hybrid OPT-mBJ calculations were performed are listed with the OPT-mBJ listed first, followed by the smaller OPT band gap in parentheses. Where the OPT-mBJ-computed band gap was found to be unrealistically small, compared to the basic OPT-calculated one, only the OPT-computed band gap is listed. Where multiple AFLOW repository entries were found to exist for the same phase of a compound, the average of each computed band gap size is listed. Literature band gaps listed in bold correspond to those experimentally-derived, while those not bolded correspond to reported, alternatively calculated band gap widths, many of which are seen to agree with the database DFT-calculations. The presence of ***** in the SC column denotes a known single-crystal synthesis method for that particular phase. (-) in any cell corresponds to uncalculated and unreported values for that particular phase, respectively.

| | | | | **Band Gap (eV)** | | | | | |
|---|---|---|---|---|---|---|---|---|---|
| **Entry #** | **Materials ID** | **Formula** | **Space Group** | **Materials Project** | **AFLOW-ML** | **Jarvis** *OPT-mBJ (OPT)* | **AFLOW** $E_{gfit}$ ($E_{gap}$) | **Literature** | **SC** |
| 1 | mp-66 | C (diamond) | $Fd\bar{3}m$ | 4.11 | 2.67 | 5.04 (4.46) | 6.46 (4.12) | **5.47**[43] | * |
| 2 | - | $C_{60}$ | $Fm\bar{3}m$ | - | - | - | - | **1.86**[44] | * |
| 3 | mp-557031 | $S_6.S_{10}$ | C2/c | 2.05 | 2.44 | - | 2.90 (1.47) | **-** | * |
| 4 | mp-77 | α-$S_8$ | Fddd | 2.71 | insulator | - | 3.44 (1.88) | **2.61**[45] | * |
| 5 | mp-558014 | $S_{12}$ | Pnnm | 2.46 | 2.31 | - | 4.19 (2.43) | - | * |
| 6 | mp-583072 | $S_{13}$ | $P2_1/c$ | 2.44 | - | - | - | - | * |
| 7 | mp-561513 | $S_{14}$ | $P\bar{1}$ | 2.43 | 2.54 | - | 4.28 (2.50) | - | * |
| 8 | mp-555915 | α-$S_{18}$ | $P2_12_12_1$ | 2.54 | - | - | - | - | * |
| 9 | mp-558964 | $S_{20}$ | Pbcn | 2.54 | - | - | - | - | * |
| 10 | mp-14 | Se | $P3_121$ | 1.06 | 1.07 | 2.29 (0.90) | 2.23 (0.98) | **1.80**[46] | * |
| 11 | mp-147 | $Se_6$ | $R\bar{3}$ | 1.54 | 1.18 | 2.12 (1.25) | 3.00 (1.55) | **1.90**[46] | * |
| 12 | mp-570481 | γ-$Se_8$ | $P2_1/c$ | 1.66 | 1.00 | - | 3.09 (1.62) | - | * |
| 13 | mp-542605 | β-$Se_8$ | $P2_1/c$ | 1.39 | - | 1.98 (1.12) | 3.28 (1.76) | - | * |
| 14 | mp-542461 | α-$Se_8$ | $P2_1/c$ | 1.57 | 0.95 | 2.16 (1.24) | 3.31 (1.78) | **2.10**[46] | * |
| 15 | mp-149 | Si | $Fd\bar{3}m$ | 0.62 | 0.38 | 1.28 (0.73) | 1.74 (0.61) | **1.12**[47] | * |
| 16 | mp-1095269 | $Si_{24}$ | Cmcm | 0.54 | - | - | - | **1.29**[48] | |
| 17 | mp-16220 | $Si_{136}$ | $Fd\bar{3}m$ | 0.53 | - | 0.56 | 1.63 (0.53) | **1.90**[49] | |

**Table 2. The potentially viable binary host materials.** All band gaps are reported in eV. Phases reported in the JARVIS-DFT database for which both standard OPT and hybrid OPT-mBJ calculations were performed are listed with the OPT-mBJ listed first, followed by the smaller OPT band gap in parentheses. Where the OPT-mBJ-computed band gap was found to be unrealistically small, compared to the basic OPT-calculated one, only the OPT-computed band gap is listed. Where multiple AFLOW repository entries were found to exist for the same phase of a compound, the average of each computed band gap size is listed. Literature band gaps listed in bold correspond to those experimentally-derived, while those not bolded correspond to reported, alternatively calculated band gap widths, many of which are seen to agree with the database DFT-calculations. The presence of ***** in the SC column denotes a known single-crystal synthesis method for that particular phase. (-) in any cell corresponds to uncalculated and unreported values for that particular phase, respectively. Grayed text indicates materials where the band gap is considered to be near enough to that of elemental Si to be potentially viable.

| | | | | Band Gap (eV) | | | | | |
|---|---|---|---|---|---|---|---|---|---|
| **Entry #** | **Materials ID** | **Formula** | **Space Group** | **Materials Project** | **AFLOW-ML** | **Jarvis** *OPT-mBJ (OPT)* | **AFLOW** *$E_{gfit}$ ($E_{gap}$)* | **Literature** | **SC** |
| 18 | mp-1735 | $BaC_2$ | I4/mmm | 1.62 | 1.47 | 3.09 (1.79 ) | 3.09 (1.61) | 2.20[50] | * |
| 19 | mp-1342 | BaO | Fm$\bar{3}$m | 2.09 | 2.24 | 3.41 (2.05) | 3.73 (2.09) | **4.10-4.38**[51] | * |
| 20 | mp-1105 | $BaO_2$ | I4/mmm | 2.24 | 2.71 | 3.74 (2.33) | 3.92 (2.23) | - | * |
| 21 | mp-1500 | BaS | Fm$\bar{3}$m | 2.16 | 1.66 | 3.23 (2.15) | 3.81 (2.15) | **3.90-4.05**[51] | * |
| 22 | mp-684 | $BaS_2$ | C2/c | 1.58 | 1.48 | 2.59 (1.63) | 3.03 (1.57) | - | * |
| 23 | mp-239 | $BaS_3$ | P$\bar{4}2_1$m | 1.37 | 1.23 | 1.02 | 2.76 (1.37) | - | * |
| 24 | mp-1253 | BaSe | Fm$\bar{3}$m | 1.96 | 1.63 | 2.85 (1.92) | 3.54 (1.95) | **3.60-3.95**[51] | * |
| 25 | mp-7547 | $BaSe_2$ | C2/c | 0.74 | 1.05 | 0.68 | 1.75 (0.62) | - | * |
| 26 | mp-7548 | $BaSe_3$ | P$\bar{4}2_1$m | 0.93 | 0.99 | 1.93 (1.00) | 2.17 (0.93) | - | |
| 27 | mp-1477 | $BaSi_2$ | Pnma | 0.81 | metal | metal | metal | **1.15**[52] | * |
| 28 | mp-1000 | BaTe | Fm$\bar{3}$m | 1.59 | 1.23 | 2.15 (1.60) | 3.06 (1.59) | **3.10-3.65**[51] | * |
| 29 | mp-8234 | $BaTe_3$ | P$\bar{4}2_1$m | 0.81 | 0.50 | 1.29 (0.76) | 1.90 (0.74) | - | * |
| 30 | mp-1197923 | $C_{61}O_2$ | $P2_1/c$ | 1.29 | - | - | - | - | |
| 31 | mp-1203337 | $C_3S_4$ | $P2_1/c$ | 1.47 | - | - | - | - | * |
| 32 | mp-30078 | $C_3S_8$ | P$\bar{3}$ | 1.73 | 2.37 | - | 3.21 (1.70) | - | * |
| 33 | mp-27814 | $CS_{14}$ | R$\bar{3}$m | 2.54 | 2.17 | 3.23 (2.28) | 4.28 (2.50) | - | * |
| 34 | mp-2482 | $CaC_2$ | I4/mmm | 1.35 | 1.90 | 2.62 (1.44) | 2.70 (1.32) | - | * |
| 35 | mp-2605 | CaO | Fm$\bar{3}$m | 3.65 | 3.16 | 5.29 (3.74) | 5.81 (3.64) | **6.10**[53] | * |
| 36 | mp-634859 | $CaO_2$ | I4/mmm | 2.73 | 3.24 | 4.22 (2.98) | 4.58 (2.72) | 2.50[54] | |
| 37 | mp-1672 | CaS | Fm$\bar{3}$m | 2.39 | 2.29 | 3.27 (2.35) | 4.13 (2.39) | **5.38**[51] | * |
| 38 | mp-1415 | CaSe | Fm$\bar{3}$m | 2.09 | 2.63 | 2.88 (2.14) | 3.71 (2.08) | **4.87-5.11**[51] | * |
| 39 | mp-1519 | CaTe | Fm$\bar{3}$m | 1.55 | 1.24 | 2.06 (1.56) | 2.99 (1.54) | **4.07-4.55**[51] | |
| 40 | mp-470 | $GeO_2$ | P4$_2$/mnm | 1.23 | 1.31 | 3.03 (1.37) | 2.48 (1.16) | **5.56**[55] | * |
| 41 | mp-2242 | GeS | Pnma | 1.24 | 1.30 | 1.32 | 2.53 (1.20) | **1.58**[56] | * |
| 42 | mp-700 | GeSe | Pnma | 0.90 | 0.75 | - | 1.98 (0.79) | **1.10**[57] | * |
| 43 | mp-10074 | $GeSe_2$ | I$\bar{4}$2d | 1.52 | 1.13 | 2.06 (1.24) | 2.96 (1.51) | **2.49**[58] | * |
| 44 | mp-352 | $HfO_2$ | P2$_1$/c | 4.02 | 3.93 | 5.66 (4.12) | 6.33 (4.02) | **5.60**[59] | * |
| 45 | mp-985829 | $HfS_2$ | P$\bar{3}$m1 | 1.24 | 1.16 | 1.68 (1.08) | 2.57 (1.23) | **2.85**[60] | * |
| 46 | mp-9922 | $HfS_3$ | P2$_1$/m | 1.12 | 1.10 | 1.93 (1.18) | 2.41 (1.11) | **3.10**[61] | * |
| 47 | mp-29771 | $MgC_2$ | P4$_2$/mnm | 2.57 | 2.15 | 3.39 (2.75) | 4.36 (2.56) | 2.64[62] | |
| 48 | mp-28793 | $Mg_2C_3$ | Pnnm | 1.65 | 2.12 | 2.41 (1.74) | 3.16 (1.67) | 2.09[63] | |
| 49 | mp-1265 | MgO | Fm$\bar{3}$m | 4.45 | 3.34 | 6.80 (5.13) | 6.94 (4.47) | **7.60**[51] | * |
| 50 | mp-1315 | MgS | Fm$\bar{3}$m | 2.79 | 3.04 | 4.26 (2.94) | 4.65 (2.77) | 3.0+[64] | * |
| 51 | mp-10760 | MgSe | Fm$\bar{3}$m | 1.77 | 2.16 | 3.37 (2.12) | 3.30 (1.77) | **5.60**[51] | |
| 52 | mp-1103590 | $MgSe_2$ | Pa$\bar{3}$ | 1.63 | 1.57 | - | 3.12 (1.64) | 3.62[65] | |
| 53 | mp-20589 | $MoO_3$ | Pnma | 1.95 | 1.57 | 2.46 (1.92 ) | 3.37 (1.82) | **3.14**[66] | * |
| 54 | mp-1018809 | $MoS_2$ | P6$_3$/mmc | 1.34 | 1.02 | 1.7 (1.44) | 2.52 (1.20) | **1.30**[67] | * |
| 55 | mp-1634 | $MoSe_2$ | P6$_3$/mmc | 1.46 | 0.85 | 1.62 (1.49) | 2.41 (1.12) | 1.41[68] | * |
| 56 | mp-19921 | PbO | P4/nmm | 1.53 | 0.95 | 2.39 (1.47) | 3.10 (1.62) | **1.90**[69] | * |
| 57 | mp-608439 | $Pb_2O_3$ | P2$_1$/m | 1.07 | 0.81 | 1.09 | 2.31 (1.04) | **1.70**[70] | |
| 58 | mp-22633 | $Pb_3O_4$ | P4$_2$/mbc | 1.14 | 1.16 | 1.73 (1.10) | 2.43 (1.13) | **2.10-2.20**[70] | * |
| 59 | mp-1285 | $PtO_2$ | Pnnm | 0.60 | 0.95 | - | 2.08 (0.68) | - | * |
| 60 | mp-2030 | $RuS_2$ | Pa$\bar{3}$ | 0.67 | 1.08 | 1.11 (0.60) | 2.63 (1.28) | **1.38**[71] | * |
| 61 | mp-726 | $SeO_2$ | P4$_2$/mbc | 3.28 | 2.89 | 3.42 | 5.29 (3.25) | - | * |
| 62 | mp-27519 | $SeO_3$ | P$\bar{4}$2$_1$c | 2.43 | - | - | 4.16 (2.41) | - | |
| 63 | mp-27358 | $Se_2O_5$ | P2$_1$/c | 2.89 | 2.96 | - | 4.81 (2.89) | - | * |
| 64 | mp-8062 | SiC (3C) | F$\bar{4}$3m | 1.39 | - | - | - | **2.39**[72] | * |
| 65 | mp-640917 | $SiO_2$ | P3$_2$21 | 0.66 | - | - | - | **9.65**[73] | * |
| 66 | mp-1602 | $SiS_2$ | Ibam | 3.07 | 2.58 | 3.56 (2.73) | 5.00 (3.03) | 2.44[74] | * |
| 67 | mp-568264 | $SiSe_2$ | Ibam | 2.16 | 1.92 | 2.51 (1.61) | 3.64 (2.02) | 3.35[74] | * |
| 68 | mp-856 | $SnO_2$ | P4$_2$/mnm | 0.65 | - | 2.38 (0.90) | 0.95 (0.03) | **3.60**[75] | * |
| 69 | mp-2231 | SnS | Pnma | 0.95 | - | 1.47 (1.02) | 2.16 (0.93) | 1.11[76] | * |
| 70 | mp-1509 | $Sn_2S_3$ | Pnma | 0.78 | 0.92 | - | 1.81 (0.67) | 1.09[76] | * |
| 71 | mp-2630 | $SrC_2$ | I4/mmm | 1.67 | 1.48 | 3.05 (1.80) | 3.17 (1.67 ) | - | |
| 72 | mp-2472 | SrO | $Fm\bar{3}m$ | 3.29 | 2.94 | 4.96 (3.40) | 5.33 (3.28) | **5.77**[51] | * |
| 73 | mp-2697 | $SrO_2$ | I4/mmm | 2.86 | 3.04 | 4.25 (3.10) | 4.71 (2.82) | - | |
| 74 | mp-1087 | SrS | $Fm\bar{3}m$ | 2.50 | 2.14 | 3.61 (2.54) | 4.28 (2.50) | **4.32**[51] | * |
| 75 | mp-1950 | $SrS_2$ | I4/mcm | 1.29 | 1.81 | 2.72 (1.58) | 2.63 (1.28) | - | * |
| 76 | mp-2758 | SrSe | $Fm\bar{3}m$ | 2.23 | 1.76 | 3.12 (2.24) | 3.92 (2.23) | **3.81**[51] | * |
| 77 | mp-1958 | SrTe | $Fm\bar{3}m$ | 1.77 | 1.18 | 2.31 (1.77) | 3.29 (1.76) | **2.90-3.40**[77] | * |
| 78 | mp-2125 | β-$TeO_2$ | Pbca | 2.23 | 2.80 | 3.76 (2.18) | 3.92 (2.23) | 2.20[78] | |
| 79 | mp-557 | α-$TeO_2$ | $P4_12_12$ | 2.81 | 2.86 | 4.64 (3.04) | 4.68 (2.81) | **3.50**[79] | * |
| 80 | mp-2552 | $TeO_3$ | $R\bar{3}c$ | 1.15 | 2.06 | - | 2.45 (1.14) | **3.25**[80] | |
| 81 | mp-390 | $TiO_2$ (anatase) | $I4_1/amd$ | 2.06 | 2.53 | 2.47 (2.02) | 4.37 (2.56) | **3.42**[81] | * |
| 82 | mp-2657 | $TiO_2$ (rutile) | $P4_2/mnm$ | 1.78 | 1.91 | 2.07 (1.77) | 3.97 (2.27) | **3.00**[82] | * |
| 83 | mp-619461 | γ-$WO_3$ | $P2_1/c$ | 1.37 | - | - | 2.76 (1.37) | **2.60-3.2**[83] | * |
| 84 | mp-1197857 | δ-$WO_3$ | $P\bar{1}$ | 1.67 | - | - | 2.94 (1.50) | - | * |
| 85 | mp-1821 | $WSe_2$ | $P6_3/mmc$ | 1.49 | 0.87 | 1.44 (1.05) | 2.54 (1.21) | **1.35**[84] | * |
| 86 | mp-1986 | ZnO | $F\bar{4}3m$ | 0.63 | 1.88 | 2.36 (0.87) | 3.19 (1.69) | **3.20**[85] | * |
| 87 | mp-8484 | $ZnO_2$ | $Pa\bar{3}$ | 2.13 | 2.71 | 3.76 (2.38) | 4.62 (2.75) | **3.80**[85] | |
| 88 | mp-10695 | ZnS | $F\bar{4}3m$ | 2.02 | 2.69 | 3.59 (2.09) | 4.52 (2.68) | **3.60**[86] | * |
| 89 | mp-1102743 | $ZnS_2$ | $Pa\bar{3}$ | 1.43 | 2.05 | - | - | **2.50**[87] | |
| 90 | mp-1190 | ZnSe | $F\bar{4}3m$ | 1.17 | 1.44 | 2.92 (1.27) | 3.20 (1.70) | **2.80**[88] | * |
| 91 | mp-2176 | ZnTe | $F\bar{4}3m$ | 1.08 | 1.25 | 2.23 (1.06) | 2.90 (1.47) | **2.40**[88] | * |
| 92 | mp-2858 | $ZrO_2$ | $P2_1/c$ | 3.47 | 3.61 | 4.21 (3.62) | 5.59 (3.47) | **5.00-6.00**[89] | * |
| 93 | mp-1186 | $ZrS_2$ | $P\bar{3}m1$ | 1.05 | 0.85 | 1.17 (0.91) | 2.32 (1.01) | **1.40**[90] | * |
| 94 | mp-9921 | $ZrS_3$ | $P2_1/m$ | 1.10 | 0.93 | 1.56 (1.21) | 2.38 (1.09) | **1.91**[91] | * |

**Table 3. The potentially viable ternary host materials.** All band gaps are reported in eV. Phases reported in the JARVIS-DFT database for which both standard OPT and hybrid OPT-mBJ calculations were performed are listed with the OPT-mBJ listed first, followed by the smaller OPT band gap in parentheses. Where the OPT-mBJ-computed band gap was found to be unrealistically small, compared to the basic OPT-calculated one, only the OPT-computed band gap is listed. Where multiple AFLOW repository entries were found to exist for the same phase of a compound, the average of each computed band gap size is listed. Literature band gaps listed in bold correspond to those experimentally-derived, while those not bolded correspond to reported, alternatively calculated band gap widths, many of which are seen to agree with the database DFT-calculations. The presence of ***** in the SC column denotes a known single-crystal synthesis method for that particular phase. (-) in any cell corresponds to uncalculated and unreported values for that particular phase, respectively. Grayed text indicates shows materials where the band gap is considered to be near enough to that of elemental Si to be potentially viable.

| | | | | **Band Gap (eV)** | | | | | |
|---|---|---|---|---|---|---|---|---|---|
| **Entry #** | **Materials ID** | **Formula** | **Space Group** | **Materials Project** | **AFLOW-ML** | **Jarvis** *OPT-mBJ (OPT)* | **AFLOW** $E_{gfit}$ $(E_{gap})$ | **Literature** | **SC** |
| 95 | mp-5504 | $BaCO_3$ | Pnma | 4.38 | - | - | - | - | * |
| 96 | - | $BaC_2O_4$ | P$\bar{1}$ | - | 3.43 | - | 5.29 (3.25) | - | * |
| 97 | mp-30237 | $Ba(CO)_4$ | I4/mcm | 2.61 | 2.95 | 3.87 (2.67) | 4.37 (2.56) | - | * |
| 98 | mp-18571 | $BaCrO_4$ | Pnma | 2.96 | 2.76 | - | 4.86 (2.93) | - | * |
| 99 | mp-572430 | $BaCr_2O_7$ | C2/c | 2.63 | - | - | - | - | |
| 100 | mp-3848 | $BaGe_4O_9$ | P321 | 3.05 | - | - | 4.97 (3.01) | - | * |
| 101 | mp-1195469 | $BaGeS_3$ | P2$_1$/c | 2.40 | - | - | - | - | * |
| 102 | mp-28710 | $BaGe_2S_5$ | Fd$\bar{3}$m | 2.19 | - | 2.09 | 3.83 (2.16) | - | * |
| 103 | mp-540805 | $Ba_2GeS_4$ | Pnma | 2.29 | 2.37 | - | 3.93 (2.24) | 2.39[92] | * |
| 104 | mp-1200759 | $Ba_3GeS_5$ | Pnma | 2.04 | - | - | - | **3.00**[93] | * |
| 105 | mp-11902 | $Ba_2GeSe_4$ | P2$_1$/m | 1.40 | 1.45 | 1.26 | 2.77 (1.38) | 3.00[94] | * |
| 106 | mp-18335 | $Ba_2Ge_2Se_5$ | Pnma | 1.29 | 1.36 | - | 2.62 (1.27) | 1.60[95] | * |
| 107 | mp-1105549 | $BaHfS_3$ | Pnma | 1.24 | - | 1.78 (0.86) | - | **2.32**[96] | |
| 108 | mp-9321 | $Ba_2HfS_4$ | I4/mmm | 0.88 | 1.38 | 1.43 (0.91) | 2.08 (0.87) | 1.58[97] | * |
| 109 | mp-1189858 | $Ba_4Hf_3S_{10}$ | I4/mmm | 0.74 | 0.41 | - | metal | - | * |
| 110 | mp-557032 | $Ba_5Hf_4S_{13}$ | I4/mmm | 0.64 | 0.45 | 0.61 | - | - | * |
| 111 | mp-554688 | $Ba_6Hf_5S_{16}$ | I4/mmm | 0.60 | 0.37 | 0.67 | - | - | * |
| 112 | mp-19276 | $BaMoO_4$ | I4$_1$/a | 4.15 | 3.13 | - | 6.24 (3.95) | - | * |
| 113 | mp-20098 | $Ba_2PbO_4$ | I4/mmm | 1.22 | 1.55 | - | 2.57 (1.23) | 1.70[98] | * |
| 114 | mp-4009 | $BaPdS_2$ | Cmcm | 0.38 | 1.08 | 0.16 (0.13) | 2.10 (0.88) | - | * |
| 115 | mp-28967 | $Ba(PdS_2)_2$ | $P2_1/m$ | 0.90 | 1.08 | 1.11 (0.97 ) | 2.37 (1.08) | - | * |
| 116 | mp-8727 | $Ba_4PtO_6$ | $R\bar{3}c$ | 1.80 | 2.20 | 1.84 | 3.61 (2.00) | - | |
| 117 | mp-29787 | $Ba_5Pt_2O_9$ | P321 | 1.93 | - | - | 3.85 (2.18) | - | |
| 118 | mp-29289 | $BaPt_2S_3$ | $P4_12_12$ | 0.97 | 1.39 | - | 2.64 (1.28) | - | * |
| 119 | mp-3164 | $BaSO_4$ | Pnma | 5.95 | 5.73 | 6.49 | 8.94 (5.97) | - | * |
| 120 | mp-1019557 | $BaS_4O_{13}$ | C2/c | 4.68 | - | - | - | - | |
| 121 | mp-6989 | $BaSeO_3$ | $P2_1/m$ | 3.90 | 3.52 | 3.98 | 6.14 (3.88) | - | * |
| 122 | mp-12010 | $BaSeO_4$ | Pnma | 3.70 | 3.62 | - | 5.89 (3.69) | - | * |
| 123 | mp-28536 | $BaSe_2O_5$ | $P2_1/c$ | 3.37 | 3.09 | - | 5.47 (3.38) | - | * |
| 124 | mp-7339 | $BaSiO_3$ | $P2_12_12_1$ | 4.53 | 3.91 | - | 7.02 (4.53) | - | * |
| 125 | mp-3031 | $BaSi_2O_5$ | Pnma | 4.76 | 4.48 | - | 6.85 (4.40) | - | * |
| 126 | mp-9478 | $BaSi_4O_9$ | $P\bar{6}c2$ | 4.35 | 4.31 | 4.50 | 6.78 (4.35) | - | * |
| 127 | mp-17612 | $Ba_2SiO_4$ | Pnma | 4.63 | - | - | - | - | |
| 128 | mp-29222 | $Ba_2Si_3O_8$ | $P2_1/c$ | 4.51 | - | - | 7.00 (4.52) | - | * |
| 129 | mp-7765 | $Ba_3SiO_5$ | I4/mcm | 3.54 | 3.24 | 3.46 | 5.67 (3.53) | - | |
| 130 | mp-669333 | $Ba_5Si_8O_{21}$ | C2/c | 4.50 | - | - | - | - | * |
| 131 | mp-5838 | $Ba_2SiS_4$ | Pnma | 3.00 | 2.70 | - | 4.97 (3.01) | - | * |
| 132 | mp-27805 | $Ba_3SiS_5$ | Pnma | 2.81 | 2.57 | - | 4.70 (2.81) | - | |
| 133 | mp-14447 | $Ba_2SiSe_4$ | $P2_1/m$ | 2.28 | 1.67 | 2.30 | 3.97 (2.27) | 3.95[94] | * |
| 134 | mp-14448 | $Ba_2SiTe_4$ | $P2_1/m$ | 1.10 | 0.92 | 1.09 | 2.40 (1.10) | - | * |
| 135 | mp-3163 | $BaSnO_3$ | $Pm\bar{3}m$ | 0.38 | - | - | metal | **3.10**[99] | * |
| 136 | mp-3359 | $Ba_2SnO_4$ | I4/mmm | 2.45 | 2.00 | - | 3.71 (2.07) | **3.18**[100] | |
| 137 | mp-12181 | $BaSnS_2$ | $P2_1/c$ | 1.62 | 1.38 | 1.76 | 3.17 (1.68) | 2.40[101] | * |
| 138 | mp-27802 | $BaSn_2S_3$ | $P2_1/m$ | 1.35 | 1.41 | - | 2.61 (1.26) | - | * |
| 139 | mp-1202711 | $BaSn_2S_5$ | Pccn | 1.53 | - | - | - | **2.35**[102] | * |
| 140 | mp-541832 | $Ba_2SnS_4$ | $P2_1/c$ | 1.96 | 1.90 | - | 3.18 (1.68) | - | |
| 141 | mp-556291 | $Ba_3Sn_2S_7$ | $P2_1/c$ | 2.19 | 1.84 | - | 3.55 (1.96) | - | * |
| 142 | mp-1204773 | $Ba_6Sn_7S_{20}$ | C2/c | 1.45 | - | - | - | - | * |
| 143 | mp-1192595 | $Ba_2SnSe_4$ | $P2_1/c$ | 1.26 | - | - | - | **1.90**[103] | * |
| 144 | mp-31307 | $Ba_2SnSe_5$ | $P2_12_12_1$ | 1.08 | - | - | - | 1.20[104] | * |
| 145 | mp-1196224 | $Ba_6Sn_6Se_{13}$ | $P2_12_12_1$ | 1.16 | - | - | - | **1.52**[105] | * |

| | | | | | | | | | | |
|---|---|---|---|---|---|---|---|---|---|---|
| 146 | mp-569754 | $Ba_7Sn_3Se_{13}$ | Pnma | 1.41 | - | - | - | 1.60-2.00[106] | * |
| 147 | mp-5431 | $BaTeO_3$ | $P2_1/m$ | 2.71 | - | - | - | - | * |
| 148 | mp-556021 | $BaTeO_3$ | Pnma | 3.17 | - | - | - | - | * |
| 149 | mp-27254 | $Ba(TeO_3)_2$ | Cmcm | 1.76 | 2.27 | 2.54 (1.91) | 3.26 (1.74) | 2.10[107] | |
| 150 | mp-1195118 | $BaTe_3O_7$ | $P2_1/c$ | 3.26 | - | - | - | - | * |
| 151 | mp-1195016 | $BaTe_4O_9$ | C2/c | 3.18 | - | - | - | - | * |
| 152 | mp-1078191 | $Ba_2TeO$ | P4/nmm | 1.78 | - | - | - | 2.93[108] | * |
| 153 | mp-17009 | $Ba_3Te_2O_9$ | $P6_3/mmc$ | 2.37 | 2.62 | - | 4.09 (2.36) | - | |
| 154 | mp-28102 | $Ba_3Te_4O_{11}$ | $P\bar{1}$ | 3.22 | - | - | - | - | * |
| 155 | mp-27499 | $BaTeS_3$ | Pnma | 1.91 | 1.67 | - | 3.49 (1.91) | - | * |
| 156 | mp-3175 | $BaTi_4O_9$ | Pmmn | 2.79 | 2.50 | - | 5.22 (3.20) | - | * |
| 157 | mp-27790 | $BaTi_5O_{11}$ | $P2_1/c$ | 2.70 | - | - | - | - | * |
| 158 | mp-504457 | $BaTi_6O_{13}$ | $P\bar{1}$ | 2.64 | 2.50 | - | 4.85 (2.92) | - | * |
| 159 | mp-3397 | $Ba_2TiO_4$ | $P2_1/c$ | 4.15 | 2.93 | - | - | - | * |
| 160 | mp-560731 | $Ba_2Ti_9O_{20}$ | $P\bar{1}$ | 2.74 | - | - | - | - | * |
| 161 | mp-29298 | $Ba_4Ti_{13}O_{30}$ | Cmce | 2.59 | - | - | - | - | * |
| 162 | mp-556729 | $Ba_6Ti_{17}O_{40}$ | C2/c | 2.71 | - | - | - | - | * |
| 163 | mp-17908 | $Ba_2TiS_4$ | Pnma | 2.10 | 1.64 | - | 3.75 (2.11) | - | |
| 164 | mp-9668 | $Ba_3TiS_5$ | I4/mcm | 1.45 | 1.51 | 1.38 | 2.77 (1.38) | - | |
| 165 | mp-19048 | $BaWO_4$ | $I4_1/a$ | 4.89 | 3.34 | 5.53 (1.66) | 7.23 (4.69) | **5.26**[109] | * |
| 166 | mp-28403 | $Ba_2WO_5$ | Pnma | 3.15 | 3.13 | - | 4.79 (2.87) | - | |
| 167 | mp-22086 | $Ba_3WO_6$ | $Fm\bar{3}m$ | 2.99 | 3.03 | - | 4.58 (2.72) | - | * |
| 168 | mp-18867 | $Ba_3W_2O_9$ | $R\bar{3}c$ | 3.67 | - | - | 5.56 (3.45) | - | |
| 169 | mp-4236 | $BaZnO_2$ | $P3_121$ | 2.27 | 2.48 | 3.92 (2.50) | 4.66 (2.78) | 2.20[110] | * |
| 170 | mp-17911 | $Ba_2ZnO_3$ | C2/c | 2.67 | 2.53 | - | 5.09 (3.10) | - | * |
| 171 | mp-5587 | $Ba_2ZnS_3$ | Pnma | 2.02 | 2.01 | - | 3.72 (2.08) | **3.31-3.49**[111] | * |
| 172 | mp-1190528 | $Ba_2ZnSe_3$ | Pnma | 1.62 | - | - | - | **2.75**[112] | * |
| 173 | mp-1190380 | $Ba_2ZnTe_3$ | Pnma | 1.32 | - | - | - | **2.10**[113] | * |
| 174 | mp-3834 | $BaZrO_3$ | $Pm\bar{3}m$ | 3.05 | 2.81 | 3.84 (3.25) | 5.00 (3.04) | - | |
| 175 | mp-8335 | $Ba_2ZrO_4$ | I4/mmm | 2.99 | 3.00 | - | 4.93 (2.98) | - | |
| 176 | mp-540771 | $BaZrS_3$ | Pnma | 1.00 | 0.97 | - | 2.29 (1.02) | **1.82**[97] | * |
| 177 | mp-3813 | $Ba_2ZrS_4$ | I4/mmm | 0.62 | 1.23 | 1.01 (0.67) | 1.73 (0.61) | **1.33**[97] | * |
| 178 | mp-9179 | $Ba_3Zr_2S_7$ | I4/mmm | 0.52 | - | - | 1.59 (0.51) | **1.21**[97] | * |
| 179 | mp-14883 | $Ba_4Zr_3S_{10}$ | I4/mmm | 0.47 | 0.84 | - | 1.54 (0.47) | - | * |
| 180 | mp-560839 | $C_4SeO_2$ | Pnnm | 1.43 | - | - | 2.92 (1.49) | - | |
| 181 | mp-553939 | $CaCO_3$ | Pnma | 4.70 | - | 6.44 (4.78) | 6.53 (4.17) | - | |
| 182 | mp-3953 | $CaCO_3$ | $R\bar{3}c$ | 5.00 | 4.50 | 6.90 (5.17) | 7.65 (5.00) | - | * |
| 183 | mp-558543 | $Ca(CO_2)_2$ | P2/m | 3.12 | 4.02 | - | 5.13 (3.13) | - | |
| 184 | mp-19215 | $CaCrO_4$ | $I4_1/amd$ | 2.37 | 2.10 | 2.59 (2.16) | 4.06 (2.34) | 2.16[114] | * |
| 185 | mp-8130 | $CaGeO_3$ | Pnma | 2.03 | - | 2.17 | - | - | * |
| 186 | mp-554678 | $CaGe_2O_5$ | $P\bar{1}$ | 2.46 | 2.51 | 2.60 | - | - | * |
| 187 | mp-560647 | $Ca_2GeO_4$ | Pnma | 3.60 | - | 3.71 | 5.72 (3.57) | - | * |
| 188 | mp-29273 | $Ca_2Ge_7O_{16}$ | $P\bar{4}b2$ | 2.87 | 2.79 | - | - | - | * |
| 189 | mp-774105 | $Ca_5Ge_3O_{11}$ | $P\bar{1}$ | 3.35 | 3.02 | - | - | - | * |
| 190 | mp-540773 | $Ca_2GeS_4$ | Pnma | 2.52 | 2.44 | - | 4.28 (2.49) | - | |
| 191 | mp-754853 | $CaHfO_3$ | Pnma | 4.49 | - | 4.57 | - | - | * |
| 192 | mp-558164 | $CaHf_4O_9$ | C2/c | 4.45 | - | - | 6.91 (4.45) | - | |
| 193 | mp-27221 | $Ca_2Hf_7O_{16}$ | $R\bar{3}$ | 4.31 | 3.74 | - | 6.73 (4.32) | - | |
| 194 | mp-558489 | $Ca_6Hf_{19}O_{44}$ | $R\bar{3}c$ | 4.13 | - | - | - | - | |
| 195 | mp-1190098 | $CaHfS_3$ | Pnma | 1.52 | - | - | - | 2.21[96] | |
| 196 | mp-19330 | $CaMoO_4$ | $I4_1/a$ | 3.51 | 2.93 | - | 5.42 (3.34) | - | * |
| 197 | mp-20079 | $CaPbO_3$ | Pnma | 0.93 | - | 1.62 (0.96) | - | - | |
| 198 | mp-21137 | $Ca_2PbO_4$ | Pbam | 1.48 | 1.79 | 2.6 (1.52) | 2.92 (1.49) | **3.65**[115] | * |
| 199 | mp-10299 | $Ca_4PdO_6$ | $R\bar{3}c$ | 1.79 | 1.94 | 1.98 (1.82) | 3.62 (2.03) | - | |
| 200 | mp-4784 | $CaPtO_3$ | Cmcm | 1.27 | 1.61 | 1.45 (1.36) | 3.10 (1.62) | - | |
| 201 | mp-8710 | $Ca_2Pt_3O_8$ | $R\bar{3}m$ | 1.08 | 1.52 | - | 3.09 (1.61) | - | |
| 202 | mp-8568 | $Ca_4PtO_6$ | $R\bar{3}c$ | 2.38 | 2.20 | 2.64 (2.41) | 4.51 (2.67) | - | * |
| 203 | mp-4406 | $CaSO_4$ | Cmcm | 5.94 | - | 6.47 | - | - | * |
| 204 | mp-1019581 | $CaS_3O_{10}$ | $P2_1/c$ | 4.92 | - | - | - | - | |
| 205 | mp-557997 | $CaSeO_3$ | $P2_1/c$ | 4.23 | - | 6.02 (4.36) | - | - | * |
| 206 | mp-1190360 | $CaSeO_4$ | $P2_1/c$ | 3.38 | - | - | 5.54 (3.43) | - | * |
| 207 | mp-27845 | $CaSe_2O_5$ | Pbca | 3.48 | 3.41 | - | 5.61 (3.49) | - | * |
| 208 | mp-28535 | $Ca_2Se_3O_8$ | $P\bar{1}$ | 3.42 | 3.48 | - | - | - | * |
| 209 | mp-5733 | $CaSiO_3$ | $P2_1/c$ | 4.88 | - | 5.17 | - | - | |
| 210 | mp-561086 | $CaSiO_3$ | C2/c | 4.71 | - | - | 7.27 (4.72) | - | * |
| 211 | mp-4428 | $CaSiO_3$ | $P\bar{1}$ | 4.73 | - | - | 7.29 (4.73) | - | |
| 212 | mp-556942 | $Ca_2SiO_4$ | $P2_1/c$ | 4.45 | 4.19 | - | 6.86 (4.41) | - | |
| 213 | mp-4481 | $Ca_2SiO_4$ | Pnma | 4.20 | - | - | 5.87 (3.68) | - | * |
| 214 | mp-641754 | $Ca_3SiO_5$ | $P\bar{1}$ | 3.99 | - | - | - | - | * |
| 215 | mp-3932 | $Ca_3Si_2O_7$ | $P2_1/c$ | 4.43 | - | - | - | - | |
| 216 | mp-1019570 | $Ca_8Si_5O_{18}$ | Pbcn | 4.55 | - | - | - | - | |
| 217 | - | $Ca_2SiS_4$ | Pnma | - | 2.96 | - | 5.09 (3.10) | - | |
| 218 | mp-1194183 | $Ca_2SiSe_4$ | Pnma | 2.47 | 2.00 | - | 4.15 (2.40) | - | |
| 219 | mp-4438 | $CaSnO_3$ | Pnma | 2.33 | - | - | - | 3.80-4.96[116] | * |
| 220 | mp-4190 | $CaSnO_3$ | $R\bar{3}$ | 2.92 | 2.07 | 3.11 | 4.89 (2.95) | **4.40**[117] | |
| 221 | mp-4747 | $Ca_2SnO_4$ | Pbam | 2.72 | 2.22 | 4.66 (2.89) | 4.23 (2.46) | - | |
| 222 | mp-866503 | $Ca_2SnS_4$ | Pnma | 2.26 | 2.07 | 1.74 (0.73) | 3.68 (2.05) | - | * |
| 223 | mp-12221 | $CaTeO_4$ | Pbcn | 2.25 | 2.49 | 2.33 | 3.96 (2.26) | - | |
| 224 | mp-5040 | $CaTe_2O_5$ | $P2_1/c$ | 3.37 | 2.66 | 3.29 | 5.43 (3.35) | - | * |
| 225 | mp-15511 | $CaTe_3O_8$ | C2/c | 2.04 | 2.32 | - | 3.61 (2.00) | - | * |
| 226 | mp-5279 | $Ca_3TeO_6$ | $P2_1/c$ | 3.25 | 2.44 | 4.56 (3.47) | 3.26 (1.74) | - | * |
| 227 | mp-680711 | $Ca_4Te_5O_{14}$ | Pbca | 2.86 | - | - | - | - | * |
| 228 | mp-559210 | $Ca_5Te_3O_{14}$ | Cmce | 1.70 | 2.36 | - | - | - | |
| 229 | mp-4019 | $CaTiO_3$ | Pnma | 2.32 | 2.66 | 2.82 (2.31) | 4.81 (2.89) | - | * |
| 230 | mp-19426 | $CaWO_4$ | $I4_1/a$ | 4.34 | 3.61 | - | 6.43 (4.10) | - | * |
| 231 | mp-779906 | $Ca_3WO_6$ | $P2_1/c$ | 3.55 | - | - | - | - | |
| 232 | mp-4571 | $CaZrO_3$ | Pnma | 3.83 | 3.16 | 4.05 | 6.06 (3.82) | - | * |
| 233 | mp-554190 | $CaZr_4O_9$ | C2/c | 3.84 | - | - | 6.09 (3.84) | - | * |
| 234 | mp-7781 | $CaZrS_3$ | Pnma | 1.19 | 1.40 | - | 2.51 (1.19) | **1.90**[118] | |
| 235 | mp-647812 | $Cr(CO)_6$ | Pnma | 3.46 | 2.93 | - | 5.81 (3.63) | - | * |
| 236 | mp-19146 | $CrPbO_4$ | $P2_1/c$ | 2.05 | 2.00 | - | metal | - | * |
| 237 | mp-22373 | $CrPb_2O_5$ | C2/m | 1.82 | 1.97 | - | 3.46 (1.89) | - | * |
| 238 | mp-705034 | $CrPb_5O_8$ | $P2_1/c$ | 1.88 | - | - | - | - | * |
| 239 | mp-643081 | $Fe_2(CO)_9$ | $P6_3/m$ | 2.48 | 3.05 | - | 4.42 (2.60) | - | |
| 240 | mp-553898 | $GePbO_3$ | P2/c | 2.70 | - | - | - | - | * |
| 241 | mp-583345 | $GePb_5O_7$ | Pbca | 2.17 | - | - | - | - | * |
| 242 | mp-29217 | $Ge_3PbO_7$ | Pbca | 2.65 | - | - | - | - | * |
| 243 | mp-680143 | $Ge_3Pb_{11}O_{17}$ | $P\bar{1}$ | 2.14 | - | - | - | - | |
| 244 | mp-624190 | $GePbS_3$ | $P2_1/c$ | 1.76 | 1.87 | 2.60 (1.66) | 3.27 (1.75) | **2.57**[119] | * |
| 245 | mp-560370 | $Ge(PbS_2)_2$ | $P2_1/c$ | 2.02 | 1.67 | 2.78 (1.93) | 3.60 (2.00) | - | * |
| 246 | mp-541785 | $GePdS_3$ | C2/m | 1.38 | 1.07 | - | 3.25 (1.73) | - | * |
| 247 | mp-1205339 | $Ge(SeO_3)_2$ | $P2_1/c$ | 3.67 | - | - | - | - | * |
| 248 | mp-1204060 | $Ge(SeO_3)_2$ | $Pa\bar{3}$ | 2.82 | 2.69 | - | 4.59 (2.73) | - | * |
| 249 | mp-554142 | $Ge(TeO_3)_2$ | $P2_1/c$ | 3.79 | 2.64 | - | - | - | * |
| 250 | mp-9755 | $HfGeO_4$ | $I4_1/a$ | 4.12 | 3.60 | 6.06 (4.17) | 6.39 (4.06) | - | * |
| 251 | mp-687090 | $Hf(MoO_4)_2$ | $P\bar{3}1c$ | 3.28 | - | - | - | - | * |
| 252 | mp-22734 | $HfPbO_3$ | Pbam | 2.72 | 2.70 | - | 4.58 (2.72) | 3.50[120] | * |
| 253 | mp-22147 | $HfPbS_3$ | Pnma | 1.46 | 1.23 | - | 2.84 (1.43) | - | * |
| 254 | mp-7787 | HfSO | $P2_13$ | 2.89 | 1.98 | 3.95 (2.87) | 4.78 (2.87) | 4.50[120] | * |
| 255 | mp-4609 | $HfSiO_4$ | $I4_1/amd$ | 5.30 | 4.62 | 5.45 | 8.06 (5.30) | - | * |
| 256 | mp-8725 | $HfSnS_3$ | Pnma | 1.22 | 1.17 | 1.17 | 2.58 (1.24) | **1.60**[121] | * |
| 257 | mp-18352 | $HfTe_3O_8$ | $Ia\bar{3}$ | 3.84 | 2.94 | - | 6.10 (3.85) | - | |
| 258 | mp-1204711 | $Hf(WO_4)_2$ | $P2_13$ | 3.92 | 3.29 | - | 6.06 (3.82) | - | |
| 259 | mp-5348 | $MgCO_3$ | $R\bar{3}c$ | 4.97 | 5.00 | 7.15 (5.22) | 7.59 (4.96) | - | * |
| 260 | mp-504578 | $MgCrO_4$ | C2/m | 2.37 | 2.52 | - | 4.08 (2.35) | - | |
| 261 | mp-19120 | $MgCrO_4$ | Cmcm | 2.47 | 2.58 | 2.52 (2.23) | 4.16 (2.41) | - | |
| 262 | mp-4575 | $MgGeO_3$ | Pbca | 2.90 | 3.06 | - | 4.73 (2.83) | - | * |
| 263 | mp-3051 | $Mg_2GeO_4$ | Pnma | 3.40 | 3.38 | - | 5.43 (3.35) | - | |
| 264 | mp-3904 | $Mg_2GeO_4$ | $Fd\bar{3}m$ | 3.07 | 2.85 | 5.22 (3.28) | 4.98 (3.01) | - | |
| 265 | mp-27295 | $Mg_{14}Ge_5O_{24}$ | Pbam | 3.45 | 3.53 | - | 5.48 (3.39) | - | * |
| 266 | mp-17441 | $Mg_2GeS_4$ | Pnma | 2.31 | 2.31 | - | 3.99 (2.29) | - | * |
| 267 | mp-1192564 | $Mg_2GeSe_4$ | Pnma | 1.42 | - | - | - | **2.02**[103] | * |
| 268 | mp-19047 | $MgMoO_4$ | C2/m | 3.79 | 3.11 | - | 5.78 (3.61) | - | * |
| 269 | mp-27604 | $MgMo_2O_7$ | $P2_1/c$ | 3.67 | 3.08 | - | 5.64 (3.51) | - | * |
| 270 | mp-7572 | $MgSO_4$ | Cmcm | 5.49 | 5.06 | 6.05 | 8.30 (5.48) | - | |
| 271 | mp-1020122 | $MgS_2O_7$ | $P\bar{1}$ | 5.52 | - | 8.05 (5.96) | - | - | |
| 272 | mp-12271 | $MgSeO_3$ | Pnma | 4.01 | 3.62 | 4.15 | 6.32 (4.01) | - | |
| 273 | mp-560670 | $MgSeO_4$ | Pnma | 3.13 | - | 4.78 (3.37) | - | - | |
| 274 | mp-16771 | $MgSe_2O_5$ | Pbcn | 3.03 | 3.18 | 3.10 | 4.98 (3.02) | - | |
| 275 | mp-556940 | $MgSiO_3$ | Pbca | 3.88 | 4.78 | - | 7.21 (4.67) | - | * |
| 276 | mp-2895 | $Mg_2SiO_4$ | Pnma | 4.64 | 4.42 | - | 7.18 (4.65) | - | * |
| 277 | mp-28663 | $Mg_{14}Si_5O_{24}$ | Pbam | 4.68 | - | - | - | - | |
| 278 | mp-1193963 | $Mg_2SiS_4$ | Pnma | 2.97 | 2.88 | - | 4.91 (2.97) | - | |
| 279 | mp-1192582 | $Mg_2SiSe_4$ | Pnma | 2.13 | 2.01 | - | 3.79 (2.13) | - | |
| 280 | mp-5746 | $MgTe_2O_5$ | Pbcn | 2.37 | 2.90 | - | 4.13 (2.38) | - | * |
| 281 | mp-769077 | $Mg_2Te_3O_8$ | C2/c | 3.48 | 2.69 | - | 5.58 (3.46) | - | * |
| 282 | mp-3118 | $Mg_3TeO_6$ | $R\overline{3}$ | 2.91 | 3.13 | 4.30 (3.11) | 4.81 (2.89) | - | * |
| 283 | mp-3771 | $MgTiO_3$ | $R\overline{3}$ | 3.50 | 3.36 | 4.02 (3.52) | 6.07 (3.83) | - | * |
| 284 | mp-28232 | $MgTi_2O_5$ | Cmcm | 2.72 | 3.15 | 3.22 (2.73) | 5.25 (3.22) | - | * |
| 285 | mp-18875 | $MgWO_4$ | P2/c | 3.68 | 3.17 | 4.18 (0.66) | 5.42 (3.34) | - | * |
| 286 | mp-25082 | $Mo(CO)_6$ | Pnma | 3.15 | 2.48 | - | 5.27 (3.23) | - | * |
| 287 | mp-22169 | $MoPbO_4$ | $I4_1/a$ | 2.90 | 2.32 | - | 4.64 (2.76) | - | * |
| 288 | mp-21138 | $MoPb_2O_5$ | C2/m | 2.93 | 2.41 | - | 4.78 (2.87) | - | * |
| 289 | mp-652366 | $MoPb_5O_8$ | $P2_1/c$ | 2.40 | - | - | - | - | * |
| 290 | mp-565836 | $MoSO_6$ | C2/c | 3.08 | - | - | 4.88 (2.94) | - | * |
| 291 | mp-640748 | $Os(CO)_4$ | $P2_1/c$ | 2.92 | - | - | - | - | * |
| 292 | mp-679947 | $Os(CO)_4$ | $P\overline{1}$ | 2.65 | - | - | - | - | * |
| 293 | mp-28636 | $Os_2(CO)_7$ | C2/c | 2.00 | - | - | - | - | * |
| 294 | mp-680248 | $Os_5(CO)_{16}$ | $P3_121$ | 1.99 | - | - | - | - | * |
| 295 | mp-647725 | $Os_5C_{17}O_{16}$ | $P\overline{1}$ | 2.02 | - | - | - | - | * |
| 296 | mp-680286 | $Os_5(CO)_{19}$ | $P\overline{1}$ | 2.03 | - | - | - | - | * |
| 297 | mp-648157 | $Os_6C_{19}O_{20}$ | Pnma | 1.77 | - | - | - | - | * |
| 298 | mp-19893 | $PbCO_3$ | Pnma | 3.33 | - | 4.87 (3.42) | - | - | |
| 299 | mp-22747 | $Pb(CO_2)_2$ | $P\overline{1}$ | 2.75 | 3.11 | 3.87 (2.86) | 4.68 (2.79) | - | * |
| 300 | mp-1179835 | $Pb(CO)_4$ | $I4_1/amd$ | 3.02 | 2.72 | - | 4.98 (3.01) | - | * |
| 301 | mp-505702 | $Pb_2CO_4$ | $P2_12_12_1$ | 3.41 | 2.99 | - | 5.53 (3.43) | - | * |
| 302 | mp-21190 | $PbSO_3$ | $P2_1/m$ | 3.24 | 3.13 | 5.17 (3.48) | 5.25 (3.21) | - | |
| 303 | mp-3472 | $PbSO_4$ | Pnma | 4.00 | 3.41 | - | 6.30 (4.00) | - | |
| 304 | mp-21904 | $PbS_2O_3$ | Pbca | 3.47 | - | - | 5.70 (3.55) | - | * |
| 305 | mp-21497 | $Pb_2SO_5$ | C2/m | 3.13 | 3.02 | 3.22 | 5.11 (3.11) | - | |
| 306 | mp-22023 | $Pb_3SO_6$ | $P2_1/m$ | 2.94 | 2.42 | - | 4.89 (2.95) | - | |
| 307 | mp-505603 | $Pb_5SO_8$ | $P2_1/c$ | 2.45 | - | - | - | - | * |
| 308 | mp-20716 | $PbSeO_3$ | $P2_1/m$ | 2.91 | 2.77 | 4.23 (3.05) | 4.74 (2.84) | - | * |
| 309 | mp-22342 | $PbSeO_4$ | $P2_1/c$ | 3.19 | 2.51 | - | 4.46 (2.63) | - | |
| 310 | mp-662535 | $PbSe_2O_5$ | $P2_1/c$ | 3.10 | 2.91 | - | 5.24 (3.21) | - | * |
| 311 | mp-558163 | $Pb(SeO_3)_2$ | $P2_1/c$ | 1.28 | 2.40 | 2.51 (1.36) | 2.61 (1.26) | - | * |
| 312 | mp-22410 | $PbWO_4$ | $I4_1/a$ | 3.56 | 2.93 | 3.94 (3.30) | 5.45 (3.37) | - | * |
| 313 | mp-22681 | $Pb_2WO_5$ | C2/m | 3.07 | 2.47 | - | 5.10 (3.10) | - | * |
| 314 | mp-28952 | $PdSO_4$ | C2/c | 1.06 | 1.77 | 1.13 | 3.66 (2.04) | - | * |
| 315 | mp-545482 | $PdSeO_3$ | $P2_1/m$ | 1.05 | 1.69 | 1.15 | 3.57 (1.97) | - | * |
| 316 | mp-561490 | $PdSeO_4$ | C2/c | 1.19 | 1.71 | 1.29 | 3.77 (2.12) | - | * |
| 317 | mp-4649 | $PdSe_2O_5$ | C2/c | 1.00 | - | 0.90 | - | - | * |
| 318 | mp-29332 | $Pt(PbO_2)_2$ | Pbam | 0.52 | 1.11 | - | 1.83 (0.68) | - | * |
| 319 | mp-652407 | $Ru(CO)_4$ | $P2_1/c$ | 2.59 | - | - | - | - | * |
| 320 | mp-21723 | $SiPbO_3$ | P2/c | 3.16 | - | - | - | - | * |
| 321 | mp-555786 | $Si(PbO_2)_2$ | C2/c | 2.52 | - | - | - | - | * |
| 322 | mp-21767 | $Si_2Pb_3O_7$ | $R\bar{3}c$ | 2.63 | 3.01 | - | 4.48 (2.64) | - | * |
| 323 | mp-1198381 | $Si_3Pb_{11}O_{17}$ | $P\bar{1}$ | 2.03 | - | - | - | - | |
| 324 | mp-504564 | $Si(PbS_2)_2$ | $P2_1/c$ | 2.03 | 1.66 | - | 3.89 (2.21) | - | * |
| 325 | mp-27532 | $Si(PbSe_2)_2$ | $P2_1/c$ | 1.82 | 1.26 | - | 3.23 (1.72) | - | * |
| 326 | mp-542769 | $Sn(CO_2)_2$ | C2/c | 2.61 | 2.50 | - | 4.48 (2.65) | - | * |
| 327 | mp-5045 | $SnGeS_3$ | $P2_1/c$ | 1.61 | 1.62 | 2.07 (1.29) | 2.89 (1.47) | **2.23**[122] | * |
| 328 | mp-504543 | $SnPbO_3$ | $Fd\bar{3}m$ | 0.00 | - | - | metal | 3.26[123] | |
| 329 | - | $Sn(PbO_2)_2$ | Pbam | - | 1.53 | - | - | - | |
| 330 | mp-542967 | $SnSO_4$ | Pnma | 4.13 | 3.61 | 4.06 | 6.24 (3.95) | - | * |
| 331 | mp-31004 | $Sn(SO_2)_2$ | $P2_1/c$ | 2.68 | 3.22 | - | 4.50 (2.66) | - | * |
| 332 | mp-28025 | $Sn_2SO_5$ | $P\bar{4}2_1c$ | 3.51 | 3.15 | - | 5.58 (3.46) | - | * |
| 333 | mp-555682 | $Sn(SeO_3)_2$ | $P2_1/c$ | 3.08 | 2.09 | - | 4.37 (2.57) | - | * |
| 334 | mp-556672 | $Sn(SeO_3)_3$ | $Pa\bar{3}$ | 2.59 | 2.56 | 2.88 | 3.72 (2.08) | - | * |
| 335 | mp-12231 | $SnTe_3O_8$ | $Ia\bar{3}$ | 2.93 | 2.62 | 3.34 | 4.41 (2.59) | - | |
| 336 | mp-17700 | $SnWO_4$ | Pnna | 0.92 | 2.66 | - | 3.09 (1.61) | - | * |
| 337 | mp-557633 | $Sn_2WO_5$ | $P2_1/c$ | 2.41 | - | - | - | - | * |
| 338 | mp-556980 | $Sn_3WO_6$ | C2/c | 2.20 | - | - | 3.72 (2.08) | - | * |
| 339 | mp-3822 | $SrCO_3$ | Pnma | 4.43 | 4.36 | - | 6.90 (4.44) | - | * |
| 340 | mp-510607 | $SrCrO_4$ | $P2_1/c$ | 2.85 | 2.14 | 2.66 | 2.86 (1.44) | - | * |
| 341 | mp-557093 | $SrCr_2O_7$ | $P4_2/nmc$ | 2.48 | - | - | - | - | * |
| 342 | mp-17464 | $SrGeO_3$ | C2/c | 3.39 | - | 5.54 (3.42) | 5.39 (3.32) | - | * |
| 343 | mp-9380 | $SrGe_4O_9$ | P321 | 3.07 | - | - | 4.97 (3.01) | - | * |
| 344 | mp-4578 | $Sr_2GeS_4$ | $P2_1/m$ | 2.33 | 2.32 | 2.21 | 3.99 (2.28) | - | * |
| 345 | mp-3378 | $SrHfO_3$ | Pnma | 4.17 | 3.88 | 4.28 | 6.53 (4.17) | - | |
| 346 | mp-1208643 | $SrHfS_3$ | Pnma | 1.49 | - | - | 2.93 (1.50) | 2.18[96] | |
| 347 | mp-18834 | $SrMoO_4$ | $I4_1/a$ | 3.81 | 3.29 | 4.27 (3.70) | 5.90 (3.70) | - | * |
| 348 | mp-20489 | $SrPbO_3$ | Pnma | 0.80 | 1.25 | 0.85 | 1.98 (0.79) | **1.80**[124] | * |
| 349 | mp-20944 | $Sr_2PbO_4$ | Pbam | 1.40 | 1.55 | 2.47 (1.43) | 2.81 (1.40) | - | |
| 350 | mp-29775 | $Sr_4PdO_6$ | $R\bar{3}c$ | 1.58 | 1.84 | 1.80 (1.60) | 3.34 (1.80) | - | |
| 351 | mp-4598 | $Sr_4PtO_6$ | $R\bar{3}c$ | 2.21 | 2.40 | 2.52 (2.22) | 4.20 (2.43) | - | * |
| 352 | mp-5285 | $SrSO_4$ | Pnma | 5.91 | 5.41 | 6.52 | 8.86 (5.90) | - | * |
| 353 | mp-3395 | $SrSeO_3$ | $P2_1/m$ | 3.71 | 3.71 | 5.05 (3.74) | 5.91 (3.70) | - | * |
| 354 | mp-555098 | $SrSeO_3$ | Pnma | 3.86 | 3.64 | 3.93 | 6.09 (3.84) | - | * |
| 355 | mp-4092 | $SrSeO_4$ | $P2_1/c$ | 3.51 | 3.66 | 5.16 (3.75) | 5.31 (3.26) | - | * |
| 356 | mp-28346 | $SrSe_2O_5$ | $P\bar{1}$ | 3.28 | - | - | 5.33 (3.28) | - | * |
| 357 | mp-3978 | $SrSiO_3$ | C2/c | 4.61 | 4.44 | 4.84 | 7.13 (4.62) | - | * |
| 358 | mp-18510 | $Sr_2SiO_4$ | $P2_1/c$ | 4.47 | 4.17 | 4.97 | 6.22 (3.93) | - | * |
| 359 | mp-5487 | $Sr_3SiO_5$ | P4/ncc | 3.76 | 3.74 | 6.02 (4.23) | 6.01 (3.78) | - | |
| 360 | mp-1104926 | $Sr_2SiS_4$ | $P2_1/m$ | 2.90 | 2.53 | - | 4.85 (2.92) | - | |
| 361 | mp-2879 | $SrSnO_3$ | Pnma | 1.74 | 1.52 | - | 2.84 (1.43) | **3.93**[125] | * |
| 362 | mp-1194203 | $Sr_2SnO_4$ | Pccn | 2.68 | 2.38 | - | 4.48 (2.65) | - | |
| 363 | mp-17743 | $Sr_3Sn_2O_7$ | Cmcm | 2.16 | - | - | 3.29 (1.76) | - | |
| 364 | mp-1179320 | $SrTeO_3$ | C2/c | 3.42 | - | - | - | - | * |
| 365 | mp-4274 | $SrTeO_4$ | Pbcn | 2.23 | 2.58 | 3.55 (2.32) | 3.94 (2.24) | - | * |
| 366 | mp-3252 | $SrTe_3O_8$ | $P4_2/m$ | 2.25 | 2.06 | 2.35 | 3.85 (2.18) | - | * |
| 367 | mp-1200112 | $Sr_3TeO_6$ | $P\bar{1}$ | 3.01 | - | - | - | - | * |
| 368 | mp-30026 | $Sr_3Te_4O_{11}$ | $P\bar{1}$ | 3.17 | 3.04 | - | 5.20 (3.18) | - | * |
| 369 | mp-706599 | $Sr_6Te_6O_{17}$ | C2/c | 1.78 | - | - | - | - | |
| 370 | mp-5229 | $SrTiO_3$ | $Pm\bar{3}m$ | 1.783 eV | 2.35 | 2.30 (1.81) | 4.00 (2.29) | - | * |
| 371 | mp-5532 | $Sr_2TiO_4$ | I4/mmm | 1.91 | 2.44 | - | 4.21 (2.45) | - | |
| 372 | mp-3349 | $Sr_3Ti_2O_7$ | I4/mmm | 1.84 | - | 1.89 | 4.12 (2.38) | - | |
| 373 | mp-31213 | $Sr_4Ti_3O_{10}$ | I4/mmm | 1.80 | - | 2.32 (1.84) | 4.07 (2.34) | - | |
| 374 | mp-19163 | $SrWO_4$ | $I4_1/a$ | 4.66 | 3.47 | - | 6.87 (4.42) | - | * |
| 375 | mp-772676 | $Sr_2WO_5$ | Pnma | 3.19 | - | - | - | - | |
| 376 | mp-5637 | $SrZnO_2$ | Pnma | 2.08 | - | 2.39 | - | - | * |
| 377 | mp-1190981 | $Sr_2ZnS_3$ | Pnma | 2.42 | 2.31 | - | 4.24 (2.47) | - | |
| 378 | mp-4387 | $SrZrO_3$ | Pnma | 3.62 | - | - | - | - | * |
| 379 | mp-27690 | $Sr_3Zr_2O_7$ | I4/mmm | 3.10 | 3.02 | - | - | - | |
| 380 | mp-5193 | $SrZrS_3$ | Pnma | 1.19 | 1.35 | 1.24 | 2.51 (1.18) | **2.05**[126] | * |
| 381 | mp-558760 | $SrZrS_3$ | Pnma | 0.55 | - | - | 1.68 (0.57) | **1.52**[126] | * |
| 382 | mp-19453 | $Te_2MoO_7$ | $P2_1/c$ | 2.81 | 2.16 | - | 4.65 (2.77) | - | |
| 383 | mp-543039 | $TePbO_3$ | C2/c | 3.04 | - | - | - | - | * |
| 384 | mp-1105700 | $TePb_2O_5$ | C2/c | 2.09 | 2.00 | - | 3.62 (2.01) | - | * |
| 385 | mp-1201764 | $TePb_5O_8$ | $P2_1/c$ | 2.16 | - | - | - | - | * |
| 386 | mp-21922 | $Te_3(PbO_4)_2$ | Cmcm | 3.00 | - | - | - | - | * |
| 387 | mp-1203659 | $Te_5PbO_{11}$ | C2/c | 3.10 | - | - | - | - | * |
| 388 | mp-29320 | $Te_3SeO_8$ | $P\bar{1}$ | 2.62 | 2.64 | 4.38 (2.79) | 4.47 (2.64) | - | * |
| 389 | mp-1008680 | TiGePt | $F\bar{4}3m$ | 0.89 | 0.50 | 0.94 | 2.16 (0.93) | 0.90[127] | * |
| 390 | mp-504427 | $Ti_3PbO_7$ | $P2_1/m$ | 1.98 | 2.52 | - | 4.08 (2.35) | - | |
| 391 | mp-554944 | $TiSO_5$ | Pnma | 2.11 | 2.84 | - | 4.12 (2.38) | - | |
| 392 | mp-29260 | $Ti(SeO_3)_2$ | $P2_1/c$ | 2.45 | 2.58 | 2.46 | 4.73 (2.83) | - | * |
| 393 | mp-18288 | $Ti(SnO_2)_2$ | $P4_2/mbc$ | 1.08 | 2.26 | - | 2.89 (1.47) | **1.55**[128] | * |
| 394 | mp-30847 | TiSnPt | $F\bar{4}3m$ | 0.80 | 0.43 | 0.83 | 1.78 (0.65) | - | * |
| 395 | mp-5214 | $TiTe_3O_8$ | $Ia\bar{3}$ | 2.73 | 2.80 | 2.73 | 5.05 (3.07) | - | * |
| 396 | mp-14142 | $TiZnO_3$ | $R\bar{3}$ | 2.98 | 3.05 | 3.05 | 5.64 (3.50) | - | |
| 397 | mp-542737 | $TiZn_2O_4$ | $P4_122$ | 2.31 | 2.91 | - | 5.09 (3.10) | - | * |
| 398 | mp-541695 | $W(CO)_6$ | Pnma | 2.98 | 2.77 | - | 5.09 (3.10) | - | |
| 399 | mp-1201710 | $WS_2O_9$ | $P2_1/c$ | 3.83 | 2.06 | - | 3.86 (2.19) | - | * |
| 400 | mp-9812 | $ZnCO_3$ | $R\bar{3}c$ | 3.56 | 3.88 | 5.94 (3.87) | 6.43 (4.09) | - | |
| 401 | mp-559437 | $Zn(CO_2)_2$ | $P2_1/c$ | 3.20 | - | 4.34 (3.27) | - | - | |
| 402 | mp-8285 | $ZnGeO_3$ | $R\bar{3}$ | 2.11 | 2.22 | 4.06 (2.37) | 4.50 (2.66) | - | * |
| 403 | mp-5909 | $Zn_2GeO_4$ | $R\bar{3}$ | 1.97 | - | - | 4.26 (2.48) | - | * |
| 404 | mp-27843 | $Zn_2Ge_3O_8$ | $P4_332$ | 2.34 | 2.50 | - | 4.50 (2.66) | - | |
| 405 | mp-16882 | $ZnMoO_4$ | $P\bar{1}$ | 3.54 | 3.08 | - | 5.85 (3.66) | - | * |
| 406 | mp-29461 | $Zn_3Mo_2O_9$ | $P2_1/m$ | 3.16 | 3.05 | 3.13 (3.01) | 5.39 (3.32) | - | * |
| 407 | mp-5126 | $ZnSO_4$ | Pnma | 3.82 | 4.09 | 4.25 | 6.92 (4.46) | - | * |
| 408 | mp-30986 | $Zn_3S_2O_9$ | $P2_1/m$ | 3.37 | 4.05 | - | - | - | * |
| 409 | mp-5338 | $ZnSeO_3$ | Pbca | 3.91 | 3.73 | 4.13 | 6.69 (4.29) | - | * |
| 410 | mp-4682 | $ZnSeO_3$ | Pnma | 3.45 | 3.59 | 3.66 | 5.94 (3.73) | - | * |
| 411 | mp-18373 | $ZnSe_2O_5$ | Pbcn | 3.04 | 3.38 | - | 4.92 (2.97) | - | * |
| 412 | mp-619034 | $ZnSiO_3$ | Pbca | 3.39 | - | - | 6.39 (4.06) | - | * |
| 413 | mp-562182 | $ZnSiO_3$ | C2/c | 3.48 | - | 5.36 (3.67) | 6.46 (4.12) | - | * |
| 414 | mp-3789 | $Zn_2SiO_4$ | $R\bar{3}$ | 2.75 | - | - | 5.56 (3.45) | - | * |
| 415 | mp-16819 | $ZnTeO_3$ | Pbca | 3.47 | 3.27 | - | 5.76 (3.59) | - | * |
| 416 | mp-560198 | $ZnTe_6O_{13}$ | $R\bar{3}$ | 2.97 | 2.57 | - | 4.95 (3.00) | - | * |
| 417 | mp-3502 | $Zn_2Te_3O_8$ | C2/c | 3.30 | 3.18 | - | 5.47 (3.38) | - | * |
| 418 | mp-13199 | $Zn_3TeO_6$ | C2/c | 1.06 | - | - | 3.25 (1.73) | - | * |
| 419 | mp-18918 | $ZnWO_4$ | P2/c | 3.22 | 2.81 | - | 5.05 (3.07) | - | * |
| 420 | mp-8042 | $ZrGeO_4$ | $I4_1/a$ | 3.73 | 3.63 | 4.84 (3.90) | 5.97 (3.75) | - | * |
| 421 | mp-27888 | $Zr_3GeO_8$ | $I\bar{4}2m$ | 3.90 | 3.47 | 4.78 (4.14) | 6.14 (3.88) | - | |
| 422 | mp-510456 | $Zr(MoO_4)_2$ | $P\bar{3}m1$ | 3.12 | 3.05 | 3.77 (3.17) | 5.02 (3.05) | - | * |
| 423 | mp-542903 | $ZrPbO_3$ | Pbam | 2.78 | 2.66 | - | 4.65 (2.77) | - | * |
| 424 | mp-20244 | $ZrPbS_3$ | Pnma | 1.25 | insulator | 1.27 | - | - | * |
| 425 | mp-3519 | ZrSO | $P2_13$ | 2.43 | 1.55 | 2.93 (2.45) | 4.17 (2.41) | 4.30[120] | |
| 426 | mp-4325 | $Zr(SO_4)_2$ | Pnma | 3.56 | - | 3.81 | 5.73 (3.57) | - | * |
| 427 | mp-4820 | $ZrSiO_4$ | $I4_1/amd$ | 4.45 | 3.95 | 5.19 (4.70) | 6.93 (4.46) | - | * |
| 428 | mp-17324 | $ZrSnS_3$ | Pnma | 1.00 | 0.90 | - | 2.34 (1.06) | **1.50**[121] | * |
| 429 | mp-4759 | $ZrTe_3O_8$ | $Ia\bar{3}$ | 3.67 | 2.79 | - | 5.85 (3.66) | - | |
| 430 | mp-1207388 | $Zr_3TiO_8$ | $I\bar{4}2m$ | 3.29 | - | - | - | - | |
| 431 | mp-18778 | $Zr(WO_4)_2$ | $P2_13$ | 3.61 | 3.30 | - | 5.70 (3.55) | 3.80[129] | |

**Table 4. The potentially viable higher-order host materials.** All band gaps are reported in eV. Phases reported in the JARVIS-DFT database for which both standard OPT and hybrid OPT-mBJ calculations were performed are listed with the OPT-mBJ listed first, followed by the smaller OPT band gap in parentheses. Where the OPT-mBJ-computed band gap was found to be unrealistically small, compared to the basic OPT-calculated one, only the OPT-computed band gap is listed. Where multiple AFLOW repository entries were found to exist for the same phase of a compound, the average of each computed band gap size is listed. Literature band gaps listed in bold correspond to those experimentally-derived, while those not bolded correspond to reported, alternatively calculated band gap widths, many of which are seen to agree with the database DFT-calculations. The presence of ***** in the SC column denotes a known single-crystal synthesis method for that particular phase. (-) in any cell corresponds to uncalculated and unreported values for that particular phase, respectively. Grayed text indicates materials where the band gap is considered to be near enough to that of elemental Si to be potentially viable.

| | | | | Band Gap (eV) | | | | | |
|---|---|---|---|---|---|---|---|---|---|
| Entry # | Materials ID | Formula | Space Group | Materials Project | AFLOW-ML | Jarvis *OPT-mBJ* *(OPT)* | AFLOW $E_{gfit}$ ($E_{gap}$) | Literature | SC |
| 432 | mp-556458 | $BaCa(CO_3)_2$ | P321 | 4.13 | 4.46 | - | 6.47 (4.13) | - | |
| 433 | mp-6568 | $BaCa(CO_3)_2$ | $P2_1/m$ | 4.69 | 4.44 | - | - | - | |
| 434 | mp-19403 | $Ba_2CaMoO_6$ | $Fm\bar{3}m$ | 2.26 | 2.29 | 2.73 | 3.77 (2.12) | - | |
| 435 | mp-17380 | $Ba_2Ca(PdO_2)_3$ | Fmmm | 1.29 | 1.48 | - | 3.70 (2.07) | - | * |
| 436 | mp-18216 | $BaCa_2(SiO_3)_3$ | $P\bar{1}$ | 4.57 | 4.46 | - | 7.06 (4.56) | - | |
| 437 | mp-550685 | $Ba_2CaTeO_6$ | $Fm\bar{3}m$ | 3.00 | 2.75 | 4.29 (3.25) | 4.95 (3.00) | - | |
| 438 | mp-18977 | $Ba_2CaWO_6$ | $Fm\bar{3}m$ | 3.18 | 3.03 | 0.57 | 4.76 (2.86) | - | |
| 439 | mp-1192417 | $BaGePbO_4$ | $P2_12_12_1$ | 3.45 | - | - | - | - | * |
| 440 | mp-1019548 | $BaGe(S_2O_7)_3$ | $P2_12_12_1$ | 3.95 | - | - | - | - | * |
| 441 | mp-570803 | $Ba_2Ge(TeSe)_2$ | $P2_1/m$ | 0.84 | 1.03 | - | 2.01 (0.82) | 1.00[130] | * |
| 442 | mp-1194160 | $BaHf(SiO_3)_3$ | $P\bar{6}c2$ | 4.68 | 4.45 | - | 7.24 (4.70) | - | |
| 443 | mp-1205395 | $BaMg(CO_3)_2$ | $R\bar{3}c$ | 4.21 | - | 6.22 (4.24) | - | - | * |
| 444 | mp-1190545 | $Ba_2MgGe_2O_7$ | $P\bar{4}2_1m$ | 3.61 | 3.40 | - | 5.71 (3.56) | - | * |
| 445 | mp-1019549 | $BaMg_2Si_2O_7$ | C2/c | 4.56 | - | - | - | - | |
| 446 | mp-9338 | $Ba_2MgSi_2O_7$ | $P\bar{4}2_1m$ | 4.46 | 4.47 | 4.70 | 6.93 (4.46) | - | * |
| 447 | mp-556703 | $Ba_3Mg(SiO_4)_2$ | $P\bar{3}$ | 4.69 | 3.88 | - | 7.25 (4.70) | - | |
| 448 | mp-18986 | $Ba_2MgWO_6$ | $Fm\bar{3}m$ | 3.23 | 2.97 | - | 4.84 (2.91) | - | * |
| 449 | mp-556779 | $BaMoSeO_6$ | $P2_1/c$ | 3.07 | 2.78 | - | 4.94 (2.99) | - | * |
| 450 | mp-1205352 | $BaPd(SeO_3)_2$ | $C2/c$ | 1.53 | - | - | - | **2.15**[131] | * |
| 451 | mp-1201400 | $BaSi(S_2O_7)_3$ | $Pbca$ | 5.29 | - | - | - | - | * |
| 452 | mp-18502 | $BaSi_3SnO_9$ | $P\bar{6}c2$ | 3.72 | 3.37 | 3.93 | 5.44 (3.36) | - | |
| 453 | mp-540635 | $BaSn(GeO_3)_3$ | $P\bar{6}c2$ | 2.77 | 2.58 | 2.87 | 4.55 (2.70) | - | |
| 454 | mp-13356 | $Ba_2SrTeO_6$ | $R\bar{3}$ | 3.13 | 2.98 | 4.42 (3.43) | 5.13 (3.12) | - | |
| 455 | mp-18764 | $Ba_2SrWO_6$ | $C2/m$ | 3.26 | - | 3.85 (0.59) | - | - | |
| 456 | mp-558894 | $Ba_4Ti_2PtO_{10}$ | $Cmce$ | 2.00 | - | - | - | - | * |
| 457 | mp-1198551 | $BaTi(S_2O_7)_3$ | $P2_1/c$ | 2.72 | - | - | - | - | * |
| 458 | mp-6661 | $BaTi(SiO_3)_3$ | $P\bar{6}c2$ | 3.07 | 3.85 | - | 5.70 (3.55) | - | |
| 459 | mp-1190949 | $Ba_2ZnGe_2O_7$ | $P\bar{4}2_1m$ | 3.10 | 3.19 | - | 5.45 (3.37) | - | * |
| 460 | mp-548469 | $BaZnSO$ | $Cmcm$ | 2.24 | 2.04 | 2.42 | 4.17 (2.42) | **3.90**[132] | * |
| 461 | mp-1205399 | $BaZn(SeO_3)_2$ | $P2_1/c$ | 3.77 | - | - | - | - | * |
| 462 | mp-558629 | $Ba_2ZnSi_2O_7$ | $C2/c$ | 4.00 | 4.12 | - | 6.67 (4.28) | - | * |
| 463 | mp-1078410 | $Ba_2ZnTeO_6$ | $Fm\bar{3}m$ | 0.43 | - | - | - | **3.00**[133] | |
| 464 | mp-548615 | $Ba_2ZnWO_6$ | $Fm\bar{3}m$ | 3.41 | 2.87 | - | 4.94 (2.99) | - | * |
| 465 | mp-1078457 | $Ba_2ZrTiO_6$ | $Fm\bar{3}m$ | 2.21 | - | - | - | - | |
| 466 | mp-1202124 | $Ba_6Zr(Zn_3S_7)_2$ | $I4/mcm$ | 1.76 | - | - | - | - | * |
| 467 | mp-15003 | $Ca_3HfSi_2O_9$ | $P2_1/c$ | 4.68 | 4.37 | - | 7.23 (4.68) | - | |
| 468 | mp-6459 | $CaMg(CO_3)_2$ | $R\bar{3}$ | 5.01 | 4.77 | 5.17 | 7.66 (5.00) | - | |
| 469 | mp-6524 | $CaMg_3(CO_3)_4$ | $R32$ | 4.83 | 4.71 | - | 7.40 (4.81) | - | |
| 470 | mp-558362 | $CaMgGeO_4$ | $Pnma$ | 3.62 | - | 5.85 (3.75) | - | - | * |
| 471 | mp-554094 | $CaMg_2(SO_4)_3$ | $P6_3/m$ | 5.76 | 5.33 | - | 8.50 (5.63) | - | * |
| 472 | mp-6493 | $CaMgSiO_4$ | $Pnma$ | 4.63 | - | 5.14 | - | - | |
| 473 | mp-562517 | $CaMg(SiO_3)_2$ | $C2/c$ | 4.66 | - | 7.36 (4.89) | 7.60 (4.96) | - | |
| 474 | mp-541026 | $CaMg_3Si_3O_{10}$ | $P\bar{1}$ | 3.41 | - | - | - | - | |
| 475 | mp-6094 | $Ca_2MgSi_2O_7$ | $P\bar{4}2_1m$ | 4.47 | 4.42 | 6.90 (4.79) | 6.93 (4.46) | - | * |
| 476 | mp-558079 | $Ca_3Mg(SiO_4)_2$ | $P2_1/c$ | 3.82 | - | - | - | - | * |
| 477 | mp-19324 | $Ca_2MgWO_6$ | $P2_1/c$ | 3.73 | 3.13 | - | 5.54 (3.43) | - | |
| 478 | mp-6796 | $Ca_5Si_2CO_{11}$ | $P2_1/c$ | 4.60 | - | - | - | - | |
| 479 | mp-6753 | $Ca_5Si_2C_2O_{13}$ | $P2_1/c$ | 4.79 | - | - | - | - | |
| 480 | mp-504590 | $Ca_2Si_3PbO_9$ | $P\bar{1}$ | 4.35 | 3.61 | - | 6.79 (4.36) | - | |
| 481 | mp-6607 | $Ca_5Si_2SO_{12}$ | $Pnma$ | 4.44 | - | - | 6.91 (4.45) | - | |
| 482 | mp-555508 | $Ca_{11}Si_4SO_{18}$ | $I\bar{4}m2$ | 3.53 | 3.64 | - | 5.64 (3.51) | - | |
| 483 | mp-6809 | $CaSiSnO_5$ | C2/c | 3.22 | - | 5.05 (3.51) | - | - | |
| 484 | mp-15095 | $Ca_3Si_2SnO_9$ | $P2_1/c$ | 3.44 | - | 3.66 | 5.09 (3.10) | - | |
| 485 | mp-559128 | $CaTeCO_5$ | Pbca | 3.85 | 3.35 | - | 6.10 (3.85) | - | |
| 486 | mp-1205333 | $CaTiGeO_5$ | C2/c | 2.72 | insulator | - | - | - | * |
| 487 | mp-1105917 | $CaTi_4(PdO_4)_3$ | $Im\bar{3}$ | 0.45 | - | - | - | 2.94-3.20[134] | |
| 488 | mp-6109 | $CaTiSiO_5$ | $P2_1/c$ | 2.96 | - | 2.96 | 3.94 (2.24) | - | * |
| 489 | mp-9413 | $Ca_2TiSiO_6$ | $Fm\bar{3}m$ | 2.36 | 3.25 | 2.96 | 4.64 (2.76) | - | |
| 490 | mp-1078391 | $CaZn(CO_3)_2$ | $R\bar{3}$ | 3.97 | - | - | - | - | |
| 491 | mp-12326 | $CaZn(GeO_3)_2$ | C2/c | 2.47 | 2.65 | - | 4.56 (2.70) | - | * |
| 492 | mp-18596 | $Ca_2ZnGe_2O_7$ | $P\bar{4}2_1m$ | 3.05 | 3.12 | 4.76 (3.13) | 5.32 (3.27) | - | * |
| 493 | mp-6693 | $CaZn(SiO_3)_2$ | C2/c | 4.06 | 4.29 | 4.40 | 6.97 (4.49) | - | * |
| 494 | mp-554905 | $CaZnSi_3O_8$ | $P\bar{1}$ | 4.09 | - | - | 6.83 (4.39) | - | * |
| 495 | mp-6227 | $Ca_2ZnSi_2O_7$ | $P\bar{4}2_1m$ | 4.06 | 4.23 | - | 6.76 (4.34) | - | * |
| 496 | mp-14778 | $Ca_3Zn_3(TeO_6)_2$ | $Ia\bar{3}d$ | 2.13 | 2.38 | - | 4.23 (2.46) | - | |
| 497 | mp-555961 | $Ca_3Zn_3(TeO_6)_2$ | $I4_132$ | 1.85 | 2.39 | - | 3.80 (2.15) | - | |
| 498 | mp-644295 | $CaZrGeO_5$ | $P\bar{1}$ | 3.90 | 3.01 | - | 3.13 (1.64) | - | |
| 499 | mp-560339 | $Ca_2Zr(SiO_3)_4$ | $P2_1/m$ | 4.85 | 4.60 | - | 7.45 (4.85) | - | * |
| 500 | mp-15004 | $Ca_3ZrSi_2O_9$ | $P2_1/c$ | 4.46 | 4.40 | - | 6.92 (4.46) | - | |
| 501 | mp-667369 | $Ca_2Zr_5Ti_2O_{16}$ | Pbca | 3.09 | - | - | - | - | |
| 502 | mp-647840 | $Fe_2C_8SO_{10}$ | $P2_1/c$ | 2.44 | - | - | - | - | * |
| 503 | mp-652326 | $Fe_3C_9S_2O_9$ | $P\bar{1}$ | 1.92 | - | - | 3.81 (2.15) | - | * |
| 504 | mp-654299 | $Fe_3C_9(SO_5)_2$ | $P\bar{1}$ | 2.17 | - | - | 3.95 (2.25) | - | |
| 505 | mp-648633 | $Fe_3C_{10}SO_{10}$ | $P2_1/c$ | 2.39 | - | - | - | - | |
| 506 | mp-653596 | $Fe_4Si(CO)_{16}$ | $P2_1/c$ | 2.62 | - | - | - | - | * |
| 507 | mp-653267 | $Fe_6Sn_2(CO)_{23}$ | $P\bar{1}$ | 2.13 | - | - | - | - | |
| 508 | mp-652337 | $Fe_3Te_2(CO)_9$ | $P\bar{1}$ | 1.72 | - | - | 3.43 (1.87) | - | * |
| 509 | mp-1197913 | $GePb(S_2O_7)_3$ | $P\bar{1}$ | 3.96 | - | - | - | - | * |
| 510 | mp-1198509 | $GePb_2(SeO_3)_4$ | $P2_1/c$ | 3.33 | - | - | 5.36 (3.30) | **4.10**[135] | * |
| 511 | mp-20056 | $MgPb_2WO_6$ | $Fm\bar{3}m$ | 2.86 | 2.63 | - | 4.40 (2.59) | - | * |
| 512 | mp-20216 | $MgPb_2WO_6$ | Pnma | 2.86 | - | - | - | - | |
| 513 | mp-1210722 | $MgTeMoO_6$ | $P2_12_12$ | 3.15 | - | - | - | - | * |
| 514 | mp-21309 | $MgTe(PbO_3)_2$ | $Fm\bar{3}m$ | 2.27 | 2.10 | - | 3.94 (2.24) | - | |
| 515 | mp-1196979 | $MoPbSeO_6$ | $P\bar{1}$ | 2.33 | - | - | 3.67 (2.04) | - | * |

| | | | | | | | | | | |
|---|---|---|---|---|---|---|---|---|---|---|
| 516 | mp-1204412 | $Mo_2PbSe_2O_{11}$ | $P\bar{1}$ | 2.50 | - | - | 4.13 (2.39) | - | * |
| 517 | mp-1196196 | $NiRu_5C_{17}O_{16}$ | $P\bar{1}$ | 1.92 | - | - | - | - | |
| 518 | mp-662792 | $Os_2C_6SO_6$ | $P\bar{1}$ | 2.00 | - | - | - | - | |
| 519 | mp-648289 | $Os_3C_8SO_8$ | $P2_1/c$ | 1.76 | - | - | - | - | * |
| 520 | mp-679948 | $Os_3C_8(SO_4)_2$ | $P2_1/c$ | 0.00 | - | - | - | - | * |
| 521 | mp-608431 | $Os_3C_9S_2O_9$ | $P\bar{1}$ | 2.51 | - | - | - | - | * |
| 522 | mp-621927 | $Os_3C_{10}SO_{10}$ | $P\bar{1}$ | 2.82 | 2.57 | - | 4.68 (2.80) | - | * |
| 523 | mp-649945 | $Os_4C_{13}SO_{13}$ | $P\bar{1}$ | 1.96 | - | - | - | - | * |
| 524 | mp-662796 | $Os_4C_{13}(SO_6)_2$ | $P\bar{1}$ | 1.90 | - | - | - | - | * |
| 525 | mp-648263 | $Os_4C_{13}S_2O_{13}$ | $P2_1/c$ | 2.03 | - | - | - | - | * |
| 526 | mp-648264 | $Os_6C_{17}S_2O_{17}$ | $P2_12_12_1$ | 1.80 | - | - | - | - | * |
| 527 | mp-662802 | $Os_6C_{19}SO_{19}$ | $P\bar{1}$ | 1.71 | - | - | - | - | * |
| 528 | mp-651026 | $Os_2C_6SeO_6$ | $P\bar{1}$ | 1.85 | - | - | - | - | * |
| 529 | mp-649754 | $Os_3C_9Se_2O_9$ | $P\bar{1}$ | 2.39 | - | - | - | - | * |
| 530 | mp-615529 | $Os_2Pt(CO)_{10}$ | $P\bar{1}$ | 2.59 | - | - | 4.30 (2.51) | - | * |
| 531 | mp-1196932 | $Pb_2C(SO_3)_2$ | Pnma | 2.80 | - | - | 4.59 (2.73) | - | |
| 532 | mp-650978 | $Ru_5C_{14}(SO_7)_2$ | $P2_1/c$ | 1.76 | - | - | - | - | * |
| 533 | mp-650984 | $Ru_6C_{17}S_2O_{17}$ | $P2_12_12_1$ | 0.90 | - | - | - | - | * |
| 534 | mp-650996 | $Ru_7C_{20}(SO_{10})_2$ | $P2_12_12_1$ | 1.07 | - | - | - | - | * |
| 535 | mp-629476 | $RuC_3SeO_3$ | $I\bar{4}3m$ | 2.19 | - | - | - | - | * |
| 536 | mp-1203202 | $Ru_3C_9Se_2O_9$ | $P\bar{1}$ | 2.24 | - | - | - | - | * |
| 537 | mp-19116 | $Sr_2CaMoO_6$ | $P2_1/c$ | 2.55 | - | - | - | - | |
| 538 | mp-11982 | $Sr_2CaTeO_6$ | $P2_1/c$ | 3.21 | - | - | - | - | |
| 539 | mp-1192743 | $SrGePbO_4$ | $P2_12_12_1$ | 3.49 | - | - | - | - | * |
| 540 | mp-1020704 | $Sr_2Ge(S_2O_7)_4$ | $P2_1/c$ | 3.43 | - | - | - | - | * |
| 541 | mp-10341 | $SrGeTeO_6$ | P312 | 2.96 | 2.70 | - | 4.90 (2.96) | - | |
| 542 | mp-972387 | $Sr_2MgGe_2O_7$ | $P\bar{4}2_1m$ | 3.61 | 3.38 | - | 5.70 (3.55) | - | * |
| 543 | mp-1078539 | $Sr_2MgMoO_6$ | I4/m | 2.29 | 2.44 | - | 3.90 (2.21) | - | |
| 544 | mp-6064 | $Sr_3MgPtO_6$ | $R\bar{3}c$ | 2.66 | 2.29 | - | 4.98 (3.01) | - | |
| 545 | mp-6564 | $Sr_2MgSi_2O_7$ | $P\bar{4}2_1m$ | 4.50 | 4.48 | - | 6.97 (4.49) | - | * |
| 546 | mp-554641 | $Sr_3Mg(SiO_4)_2$ | C2/c | 4.69 | - | - | - | - | |
| 547 | mp-1078995 | $Sr_2MgTeO_6$ | C2/m | 2.33 | - | - | - | - | |
| 548 | mp-19420 | $Sr_2MgWO_6$ | I4/m | 3.43 | 3.04 | - | 5.05 (3.07) | - | |
| 549 | mp-1200408 | $SrMo_2Se_2O_{11}$ | $P\bar{1}$ | 2.73 | - | - | 4.41 (2.59) | - | * |
| 550 | mp-556331 | $Sr_4Pt_4PbO_{11}$ | $P\bar{1}$ | 0.57 | 1.54 | - | 2.20 (0.96) | - | * |
| 551 | mp-1020612 | $Sr_2Si(S_2O_7)_4$ | $P2_1/c$ | 4.91 | - | - | - | - | * |
| 552 | mp-559957 | $SrTiGeO_5$ | $P2_1/c$ | 2.84 | 2.79 | - | - | - | |
| 553 | mp-559208 | $SrTiSi_2O_7$ | Cmcm | 3.61 | 3.71 | 3.73 (3.66) | 6.62 (4.24) | - | |
| 554 | mp-558553 | $Sr_4Ti_5(Si_2O_{11})_2$ | C2/m | 1.94 | - | - | - | - | |
| 555 | mp-17392 | $Sr_2ZnGe_2O_7$ | $P\bar{4}2_1m$ | 3.12 | 3.09 | - | 5.42 (3.34) | - | * |
| 556 | mp-1079971 | $Sr_2ZnMoO_6$ | I4/m | 2.38 | 2.31 | - | 4.01 (2.30) | - | |
| 557 | mp-6730 | $Sr_3ZnPtO_6$ | $R\bar{3}c$ | 2.56 | 2.27 | 2.65 | 4.93 (2.98) | - | * |
| 558 | mp-555691 | $SrZn(SeO_3)_2$ | $P2_1/c$ | 4.33 | 3.62 | - | 6.75 (4.33) | - | * |
| 559 | mp-1191240 | $Sr_2ZnSi_2O_7$ | $P\bar{4}2_1m$ | 3.90 | 4.13 | - | 6.68 (4.28) | - | |
| 560 | mp-1079778 | $Sr_2ZnWO_6$ | I4/m | 3.53 | - | - | - | - | |
| 561 | mp-1105272 | $Sr_2ZnWO_6$ | $P2_1/c$ | 3.51 | - | - | - | - | |
| 562 | mp-17468 | $SrZrSi_2O_7$ | $P2_1/c$ | 4.71 | - | - | - | - | |
| 563 | mp-555314 | $Sr_7Zr(Si_2O_7)_3$ | $R\bar{3}$ | 4.43 | - | - | - | - | * |
| 564 | mp-1080028 | $Sr_2ZrTiO_6$ | I4/m | 2.51 | - | - | - | - | |
| 565 | mp-651231 | $Te_2Ru_4(CO)_{11}$ | Pccn | 1.74 | - | - | - | - | * |
| 566 | mp-651051 | $Te_2W_3(CO)_{15}$ | $P4_12_12$ | 1.22 | - | - | - | - | * |
| 567 | mp-1203799 | $TiPb(S_2O_7)_3$ | $P2_1/c$ | 2.72 | - | - | - | - | * |
| 568 | mp-558830 | $ZnTeMoO_6$ | $P2_12_12$ | 3.09 | 2.78 | 3.70 (2.98) | 4.98 (3.02) | - | * |
| 569 | mp-1103554 | $BaCa_2Mg(SiO_4)_2$ | $P\bar{3}$ | 4.97 | 4.46 | - | 7.63 (4.99) | - | |
| 570 | mp-1019558 | $BaSr_2Mg(SiO_4)_2$ | $P\bar{3}m1$ | 4.76 | - | 6.89 (5.26) | - | - | |
| 571 | mp-17244 | $Ba_2ZnGe_2S_6O$ | $P\bar{4}2_1m$ | 2.09 | 2.30 | - | 3.72 (2.08) | - | * |
| 572 | mp-543034 | $Ba_3Zn_6Si_4TeO_{20}$ | C2/m | 2.64 | 3.08 | - | 5.29 (3.24) | - | * |
| 573 | mp-1200437 | $Ca_2Zn_3Si_3PbO_{12}$ | $P2_1/c$ | 3.71 | - | - | - | - | |
| 574 | mp-652318 | $CrFe_2C_{10}(Se_2O_5)_2$ | $P2_1/c$ | 1.52 | - | - | - | - | * |
| 575 | mp-651724 | $Fe_2Te_2Ru_2(CO)_{11}$ | Pccn | 1.70 | - | - | - | - | * |
| 576 | mp-559754 | $Fe_2Te_2W(CO)_{10}$ | $P\bar{1}$ | 1.93 | - | - | - | - | * |
| 577 | mp-557864 | $Fe_2WC_{10}(SeO_5)_2$ | $P\bar{1}$ | 1.96 | - | - | - | - | * |
| 578 | mp-558938 | $Zn_4SiTePbO_{10}$ | Pnma | 2.29 | - | - | - | - | * |
| 579 | mp-683831 | $Fe_3MoC_{11}SeS_3O_{11}$ | $P2_1/c$ | 2.12 | - | - | - | - | * |
| 580 | mp-624009 | $Fe_2TeWC_{10}SeO_{10}$ | $P\bar{1}$ | 1.91 | - | - | - | - | * |